\documentclass[%
 reprint, superscriptaddress,
 amsmath,amssymb,
 aps,
]{revtex4-2}

\usepackage{graphicx}
\usepackage{dcolumn}
\usepackage{bm}
\usepackage{bbold}

\usepackage{physics}
\usepackage{amsmath}
\usepackage{hyperref}
\hypersetup{
    colorlinks,
    citecolor=blue,
    filecolor=blue,
    linkcolor=blue,
    urlcolor=blue
}
\usepackage{natbib}

\newcommand{\ba}{\begin{eqnarray}}
\newcommand{\ea}{\end{eqnarray}}
\newcommand{\ban}{\begin{eqnarray*}}
\newcommand{\ean}{\end{eqnarray*}}

\begin{document}

\title{The role of photon-number pulses in the operation of a simple light diode}

\author{Zakarya Lasmar}
\affiliation{Centre for Quantum Technologies, National University of Singapore, 3 Science Drive 2, Singapore 117543}
\author{Shuaijie Li}
\affiliation{School of Physics, Xi’an Jiaotong University, Xi’an 710049, China}
\author{Valerio Scarani}
\affiliation{Centre for Quantum Technologies, National University of Singapore, 3 Science Drive 2, Singapore 117543}
\affiliation{Department of Physics, National University of Singapore, 2 Science Drive 3, Singapore 117542} 

\date{\today}

\begin{abstract}
One of the challenges faced by optical platforms for quantum technologies is the implementation of (ultimately) a transistor. The functionality that is hard to achieve is rectification: having the beam propagating in one direction transmitted, the other reflected. Here we take up a simple model of such a rectifying device, a.k.a.~optical diode, consisting of two atoms with different detuning interacting with light in a one-dimensional waveguide. In previous studies, it was found that high rectifying efficiencies can be achieved with coherent states, while it was claimed that the device cannot rectify single-photon Fock states. In this paper, we clarify the functioning of this diode. Notably, we show that coherences across the Fock bases in the input state do not play any role, and thence the rectifying properties of the device depend only on its behavior on the Fock states. In the process, we show that some single-photon rectification is predicted when the limit of infinitely-long pulses is not taken.

\end{abstract}

\maketitle

\section{Introduction} 
Controlling the photonic transport in integrated circuits is important for building future information and communication technologies. While a remarkable progress has been achieved on early implementations of such technologies \cite{Asavanant,Larsen,Juan}, the development of an optical rectifier that can act on single or a few photon pulses remain challenging. These rectifying devices are useful for routing and isolating signals. Rectifiers based on ferromagnetic compounds have been demonstrated for increasing signal processing capabilities. However, they are inherently lossy and difficult to miniaturize and use in integrated circuits \cite{Pozar}. Recently, using a semi-classical theory, a two atoms device has been identified as a candidate for exhibiting strong directionality \cite{auffeves2014}. The transmission of light through this device would depend on the direction of propagation. If the input comes from one side it will pass through it, and if it comes from the other side it will get reflected. This has triggered a debate over whether such a device can perform with non-classical states of light \cite{Alex,Auffeves2016,Vienna2016,Combes,Fang,Clemens,Gonzalez,Ordonez}.

This rectifying device, also referred to as \textit{optical diode}, is consisting of a pair of two-level systems coupled to a one-dimensional waveguide. By changing the distance between the atoms and the detuning of their resonance frequencies, the efficiency of the diode was optimized. It has been shown that for light pulses in coherent states, the device can exhibit efficiencies of more than 60\% \cite{Alex,Auffeves2016,Vienna2016,Combes}. However, in the monochromatic limit, it has been found that this device cannot rectify single photon pulses \cite{Alex}. Since significant directionality can be achieved for coherent inputs while a symmetric behaviour is predicted for single photon pulses, one might naturally ask: which components of the coherent state triggers the performance of the device? Do superpositions matter? How can we change these critical components, so we can improve the efficiency of this device for inputs with a few photon pulses? To answer this questions, we will use the same approach as in \cite{Alex}. Using the Heisenberg equations of motion, we will show that by changing the length of the light pulse, we can reduce the number of photons necessary to activate the rectifying behaviour of the optical diode. 

This article is organized as follow: in section II, we will introduce the theoretical model for this device and present how the rectification factor of the device is computed. In section III, we will discuss the cases of photon-number pulses. We will show that the main terms that will contribute to the rectification depend directly on the pulse duration. In section IV, we will present a few concluding remarks and discuss potential future directions of this research.

\section{The model}

\subsection{The rectifying device}

\begin{figure}
    \centering
        \includegraphics[width=0.5\textwidth,trim={0 9cm 0 0cm},clip]{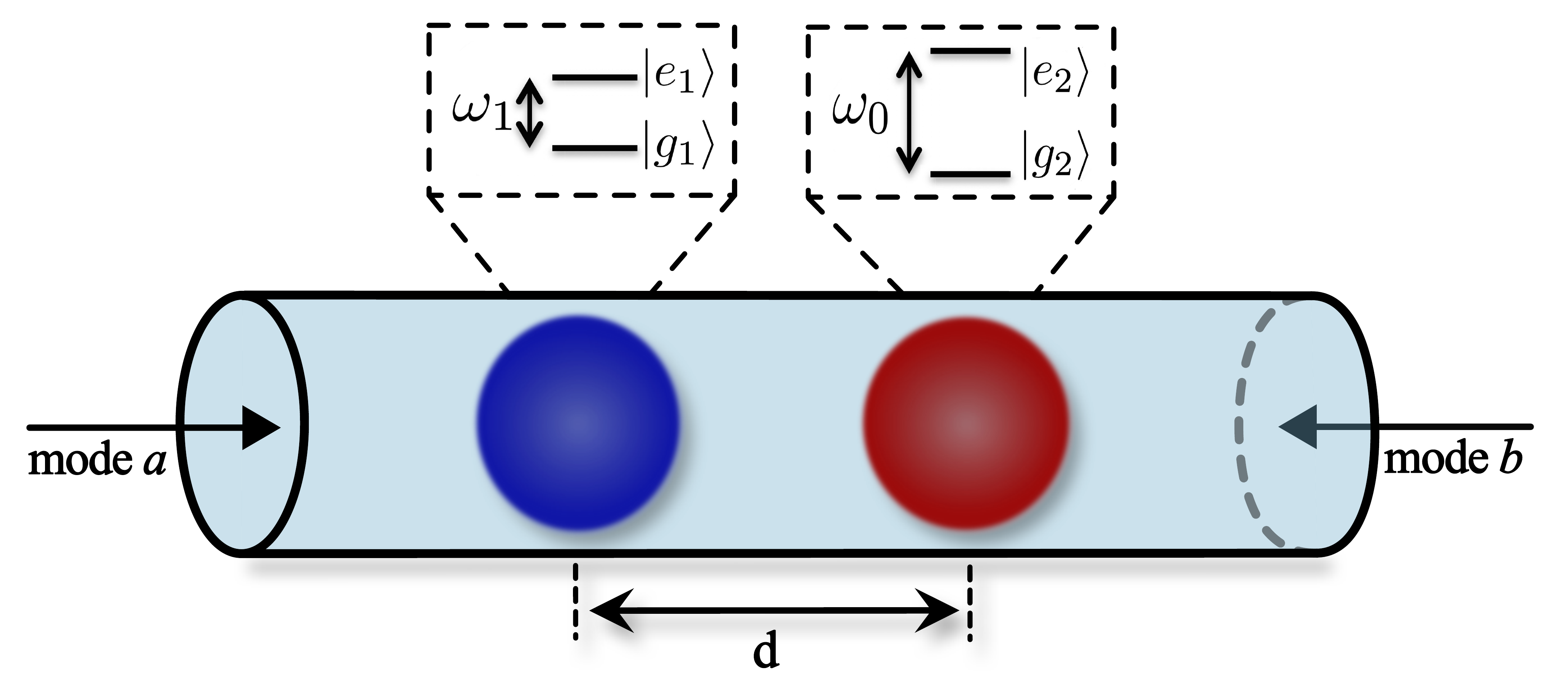}
    \caption{A scheme of the rectifying device consisting of two 2-level systems coupled to a one-dimensional waveguide. The one on the right is resonant with the central frequency of the light pulse, $\omega_0$. The atom on the left is detuned, $\omega_1 \neq \omega_0$.}
    \label{fig:diode}
\end{figure}

As illustrated in Fig.~\ref{fig:diode}, the system of interest here is composed of a pair of atoms strongly coupled to a one dimensional waveguide. Each atom is model as a two-level system, for which $\ket{g_j}$ and $\ket{e_j}$ are the ground and excited states of the $j^{th}$ atom, respectively. The atom on the right is resonant with the central frequency of the light pulse $\omega_0$. On the other hand, the atom on the left is detuned. We denote its resonance frequency by $\omega_1$. Both frequencies are assumed to be much larger than the cutoff frequency of the waveguide \cite{Snyder}. In this case, the dispersion relations read $\omega = v_g \abs{k}$, such that $k$ is the longitudinal wave-number of the field mode and $v_g$ is the group velocity of the light pulse in the waveguide. The free Hamiltonian of this system reads
\begin{equation} \label{H0}
    \hat{\mathcal{H}}_{0} = \sum_{j=1}^{2}  \hbar \omega_j \ketbra{e_j} + \int_{0}^{\infty} \dd \omega \hbar \omega  (\hat{a}^{\dagger}_{\omega}\hat{a}^{}_{\omega} + \hat{b}^{\dagger}_{\omega}\hat{b}^{}_{\omega}) \;,
\end{equation}
such that the first term corresponds to the two atoms. The second term describes the field modes. The operator $\hat{a}^{\dagger}_{\omega}$ ($\hat{b}^{\dagger}_{\omega}$) creates a photon with frequency $\omega$ moving towards the right (left). The interaction between the atoms and the propagating light pulse is described by the dipole Hamiltonian within the rotating wave approximation \cite{Snyder,Roulet}
\begin{equation} \label{Hdip}
\begin{split}
    \hat{\mathcal{H}}_{\mathrm{dip}} =& -i \hbar \sum_{j=1}^{2} \int_{0}^{\infty} \dd \omega g_{\omega}^{(j)} \\
    \times \Big[ & \hat{\sigma}_{+}^{(j)} \left( \hat{a}_{\omega} e^{i \omega \frac{ x_j}{v_g}} + \hat{b}_{\omega} e^{- i \omega \frac{ x_j}{v_g}}\right)e^{ - i (\omega - \omega_j)t} - \textbf{H.c.} \Big] \;.
\end{split}
\end{equation}
Here $x_j$, $g_{\omega}^{(j)}$ and $\hat{\sigma}_{+}^{(j)} = \ketbra{e_j}{g_j} $ are the position, the coupling strength and raising operator of the $j^{th}$ atom, respectively. Similarly to the previous study \cite{Alex}, we assume the Weisskopf-Wigner approximation \cite{Scully}, i.e. both atoms have the same coupling to the waveguide, $g_{\omega}^{(1)} = g_{\omega}^{(2)} = g$. Thus, the decay rate to the waveguide for both atoms is $\gamma = 2\pi g^2$.

\subsection{Computing rectification}

Several figures of merit have been used in previous papers to capture rectification: we review them in Appendix \ref{app:defs}. Here we define the rectifying factor of the device as
\begin{equation}\label{DiodeEff}
    \mathcal{R} = T_{\rightarrow} - T_{\leftarrow}\,,
\end{equation} where $T_{\rightarrow}$ is the transmittivity of the light coming from the left (i.e.~the fraction of left-incoming light scattered towards the right); analogously, $T_{\leftarrow}$ is the fraction of right-incoming light scattered towards the left. Clearly $-1\leq\mathcal{R}\leq 1$, the sign determining the direction in which the diode rectifies. Explicitly,
\ba\label{eqsT}
T_{\rightarrow}&=&\frac{1}{F}\lim_{t\rightarrow\infty}N_a^{\rightarrow}(t)\,=\,\frac{1}{F}(1-\lim_{t\rightarrow\infty}N_b^{\rightarrow}(t))\,,\\
T_{\leftarrow}&=&\frac{1}{F}\lim_{t\rightarrow\infty}N_b^{\leftarrow}(t)
\ea
where $F$ is the mean photon flux per unit time, and where \ban
N_{b}^{\textrm{input}} (t) = \int_{0}^{\infty} \dd \omega \textrm{Tr}\left[\hat{b}^{\dagger}_{\omega}(t)\hat{b}^{}_{\omega}(t)\,\rho_{\textrm{input}}\right]\;,
\ean (and similar definitions for similar quantities).

In this paper, we shall consider monomode Fourier-limited pulses. The mode that defines a pulse coming from the left is 
\begin{equation}\label{hatA}
\hat{A}^\dagger = \int_{0}^{\infty} \dd\omega f (\omega) \hat{a}_{\omega}^{\dagger} = \int_{0}^{\infty} \dd\tau \xi (\tau) \hat{a}_{\tau}^{\dagger}  \;,
\end{equation}
where
\begin{equation}
    \begin{split}
    \hat{a}_{\tau}^{} =& \frac{1}{\sqrt{2\pi}} \int \dd\omega \hat{a}_{\omega}^{} e^{- i (\omega - \omega_0) \tau} \;,\\
    \xi (\tau) =& \frac{1}{\sqrt{2\pi}} \int \dd\omega f (\omega) e^{- i (\omega - \omega_0) \tau}\,.
    \end{split}
\end{equation} For pulses coming from the right, we use the identical definition of a mode $\hat{B}$, replacing the operators $(\hat{a}_{.},\hat{a}^\dagger_{.})$ with $(\hat{b}_{.},\hat{b}^\dagger_{.})$.

Like in previous works, for definiteness the calculations will assume a square pulse
\begin{equation}\label{xitau}
\xi (\tau) = \Bigg\{
    \begin{array}{cl}
        \sqrt{\Omega/2}  &  \text{for } 0 \leq \tau \leq 2/\Omega,  \\
        0 & \text{otherwise}.
    \end{array}
\end{equation}
For an input state $\ket{\psi}$, the mean flux of photons coming from the left is 
\begin{equation}\label{flux}
    F = \textrm{Tr}\left[\hat{a}_\tau^\dagger(0)\hat{a}_\tau(0)\rho_{\textrm{input}}\right]\;,
\end{equation}

A basis of the total states of the system is denoted $\ket{n_a,n_b,s_1,s_2}$ where $n_{a,b}$ is a Fock state in mode $\hat{A}$ or $\hat{B}$, while $s_j\in\{g,e\}$ is the state of the $j$-th atom. Initially, the pulse will come from either the left or the right; and both atoms are assumed to be in the ground state to ensure that the rectifying device is passive. Thus our input states will always be of the form $\ket{\psi_a}\equiv\ket{\psi,0,g,g}$, or $\ket{\psi_b}\equiv\ket{0,\psi,g,g}$.

Instead of solving the time-dependent equations of motion and sending $t\rightarrow\infty$, we shall consider only long pulses and thus assume that the system is in the steady state:
\ba
\lim_{t\rightarrow\infty}N_b^{\textrm{input}}(t)&\approx&\int_{0}^{\infty} \dd \omega \expval{\hat{b}^{\dagger}_{\omega}\hat{b}^{}_{\omega}}_{ss,\textrm{input}}\,.\label{ss:assume}
\ea

\subsection{The case for pulses of finite length}

All previous fully-quantum studies of this rectifying device \cite{Alex,Vienna2016,Auffeves2016} concluded that there is no rectification for single-photon Fock states. This conclusion was drawn in the limit of pulses of \textit{infinite length}. One of the results of the next Section will be that \textit{the device does show some rectification for single-photon Fock states} for pulses of finite length.

Those same works also showed that high rectification is expected for coherent states, in qualitative (if not exact quantitative) agreement with a previous semi-classical study \cite{auffeves2014}. Specifically, rectification was predicted \cite{Alex} and later observed \cite{Clemens} to have a high-value plateau when $F/\gamma$ ranges between $10^{-3}$ and $10^{-1}$. Let us now describe this case in more detail, to understand what may be the origin of the rectification.

Every coherent state is ultimately monomode \cite{Alex90s,Alex90ss}: a coherent state pulse with average number of photons $\bar{n}$, propagating towards the right, can be written as
\begin{equation}\label{state:cohbis}
    \ket{\alpha_a} = \exp\big(\sqrt{\bar{n}} (\hat{A}^\dagger - \hat{A})\big) \ket{\text{\o{}}}\;.
\end{equation} Since $\hat{a}_\tau(0)\ket{\alpha} = \sqrt{\bar{n}}\, \xi(\tau) \ket{\alpha}$, the flux \eqref{flux} is given by $F=\frac{\bar{n} \Omega}{2}$. If the device performance is maximized for plateau centred around $F/\gamma \approx  10^{-2}$, then the optimal average number of photons in the pulse is
\begin{equation}
   \bar{n}_{\textrm{opt}} \approx \frac{2 \gamma }{\Omega}  \times 10^{-2}\,.
\end{equation}
If we were to consider infinitely long pulses, as used in the single-photon case, then we would have to take the monochromatic limit $\Omega \rightarrow 0$; in which case $\bar{n}_{\textrm{opt}}$ diverges.

\begin{figure}
    \centering
        \includegraphics[width=0.23\textwidth]{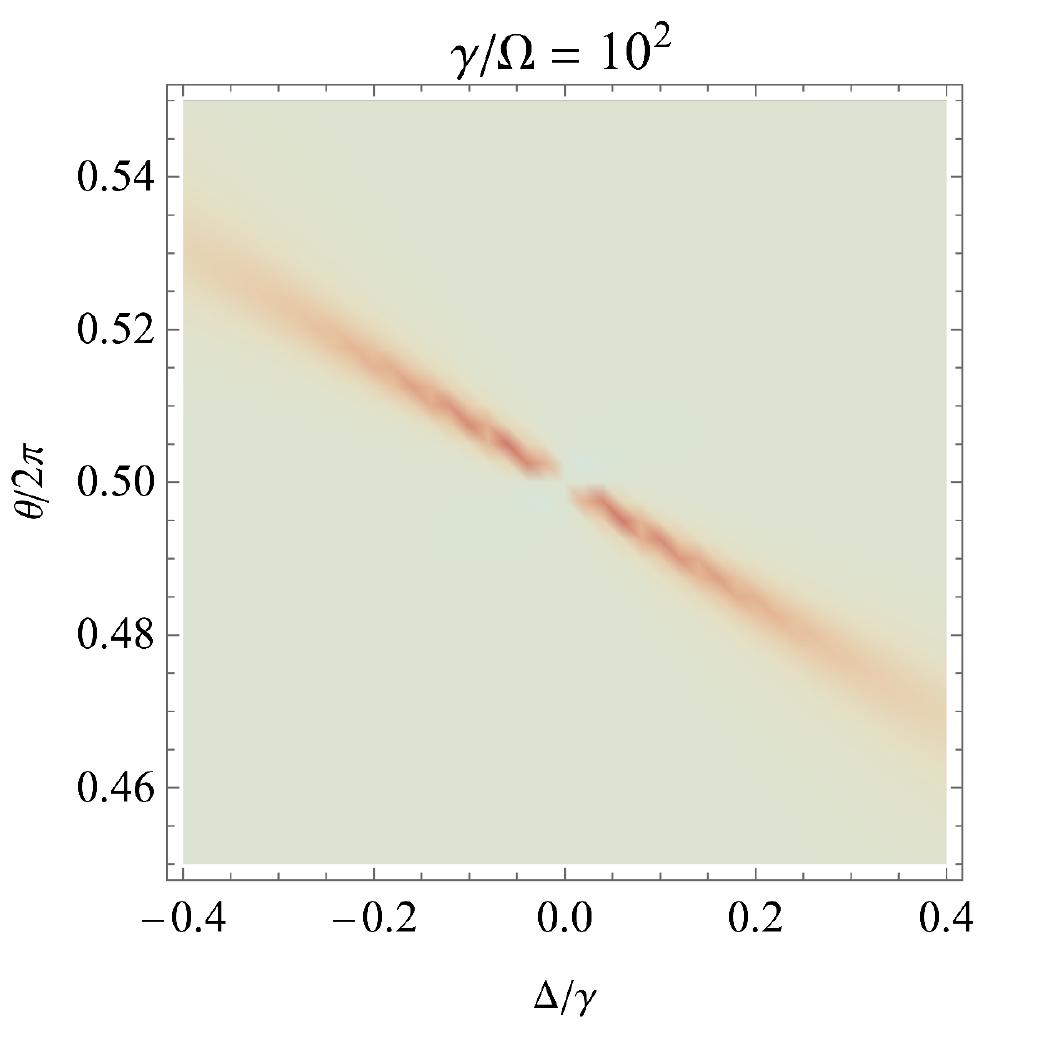}
        \includegraphics[width=0.23\textwidth]{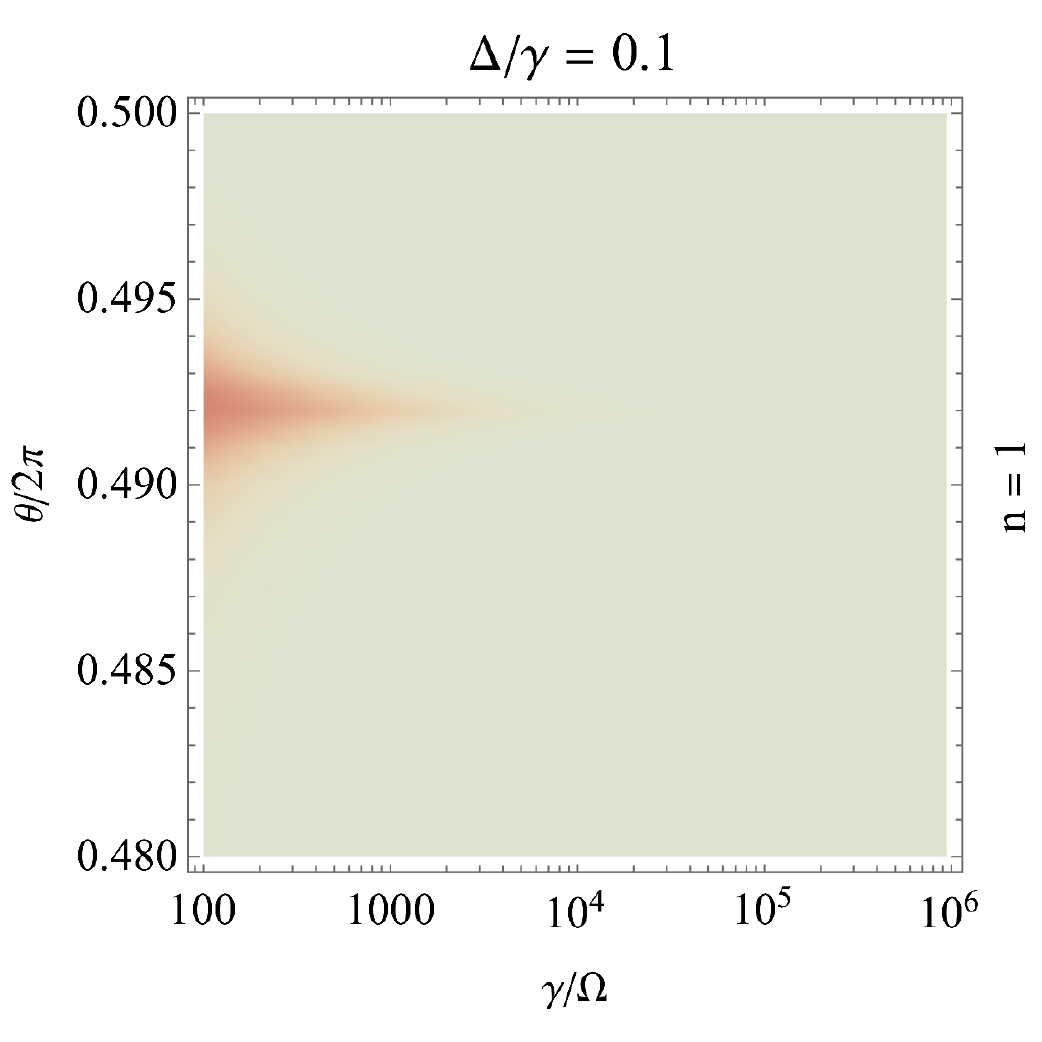}\\
        \includegraphics[width=0.23\textwidth]{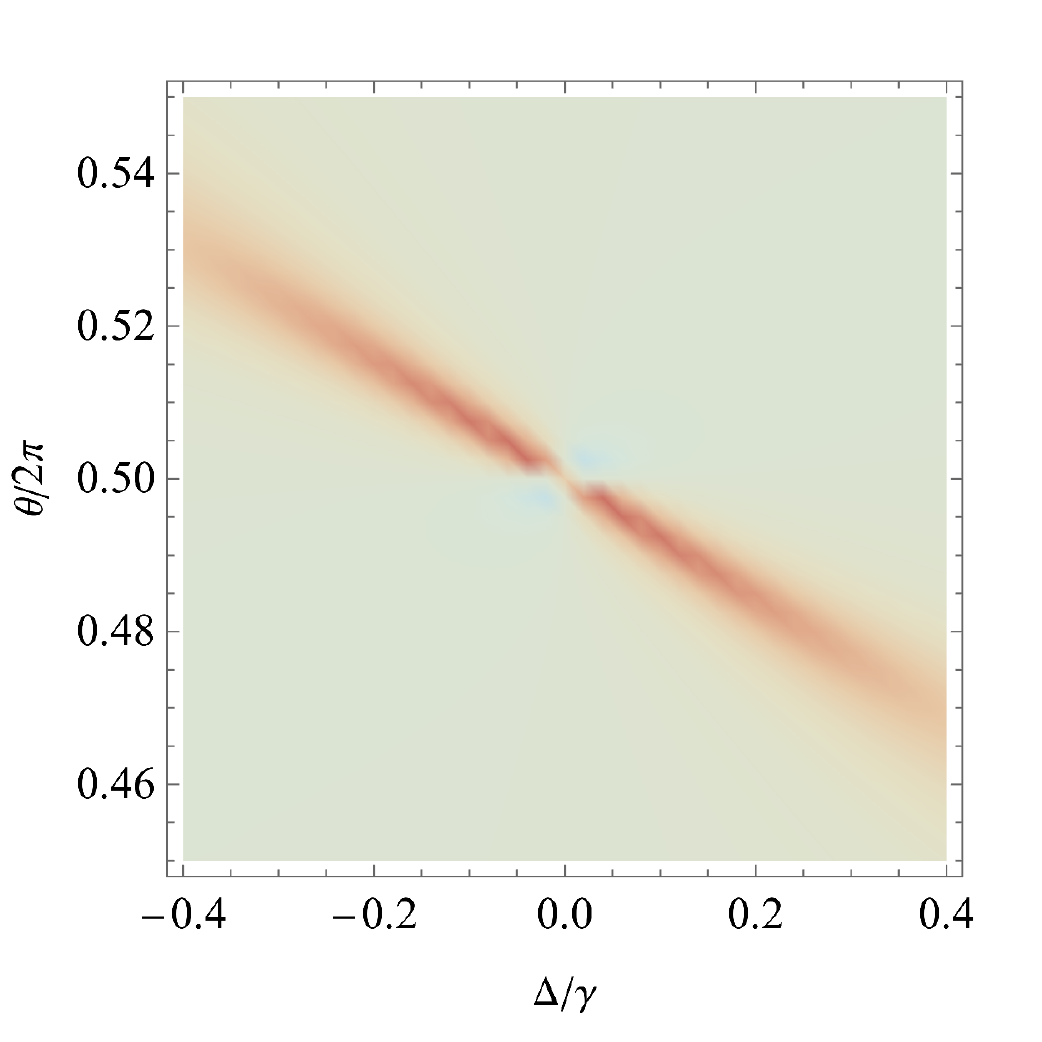}
        \includegraphics[width=0.23\textwidth]{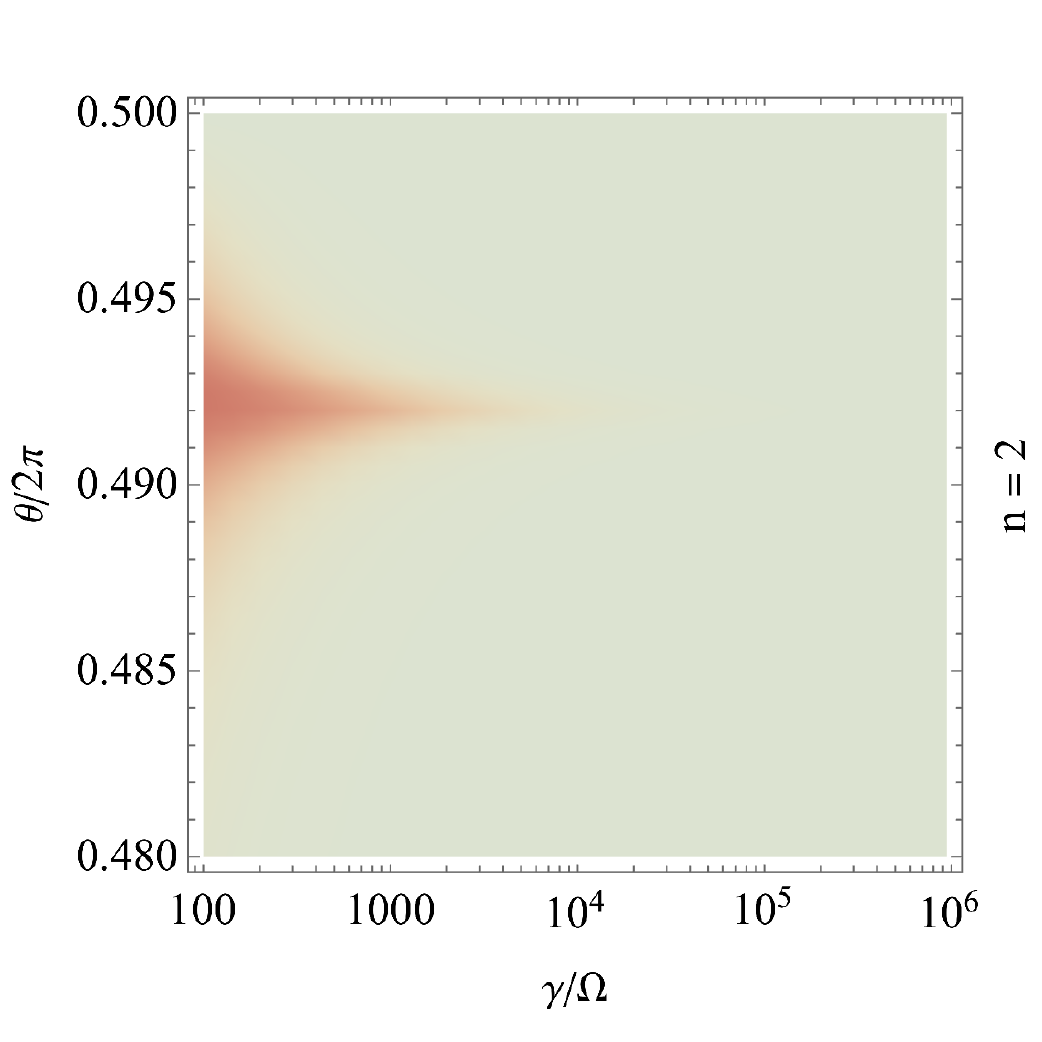}\\
        \includegraphics[width=0.23\textwidth]{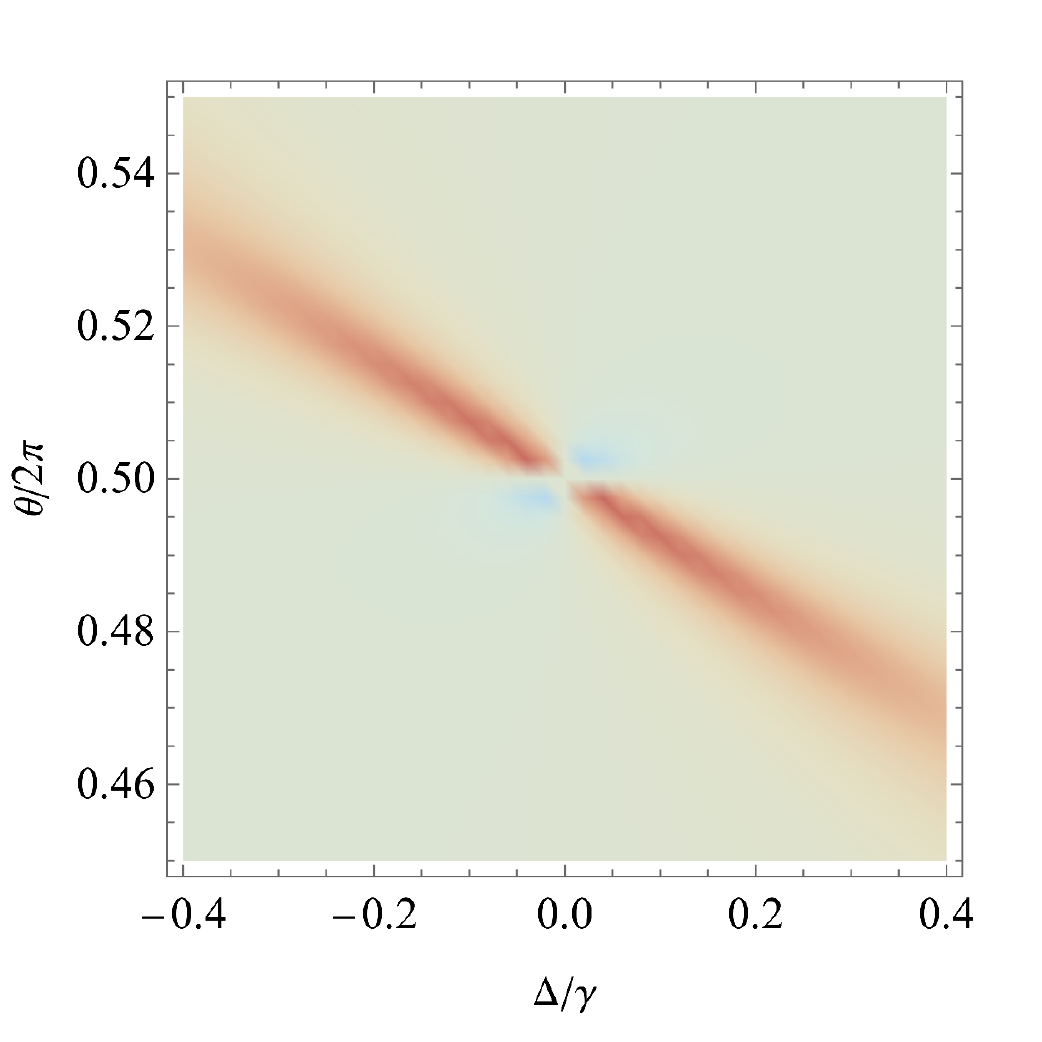}
        \includegraphics[width=0.23\textwidth]{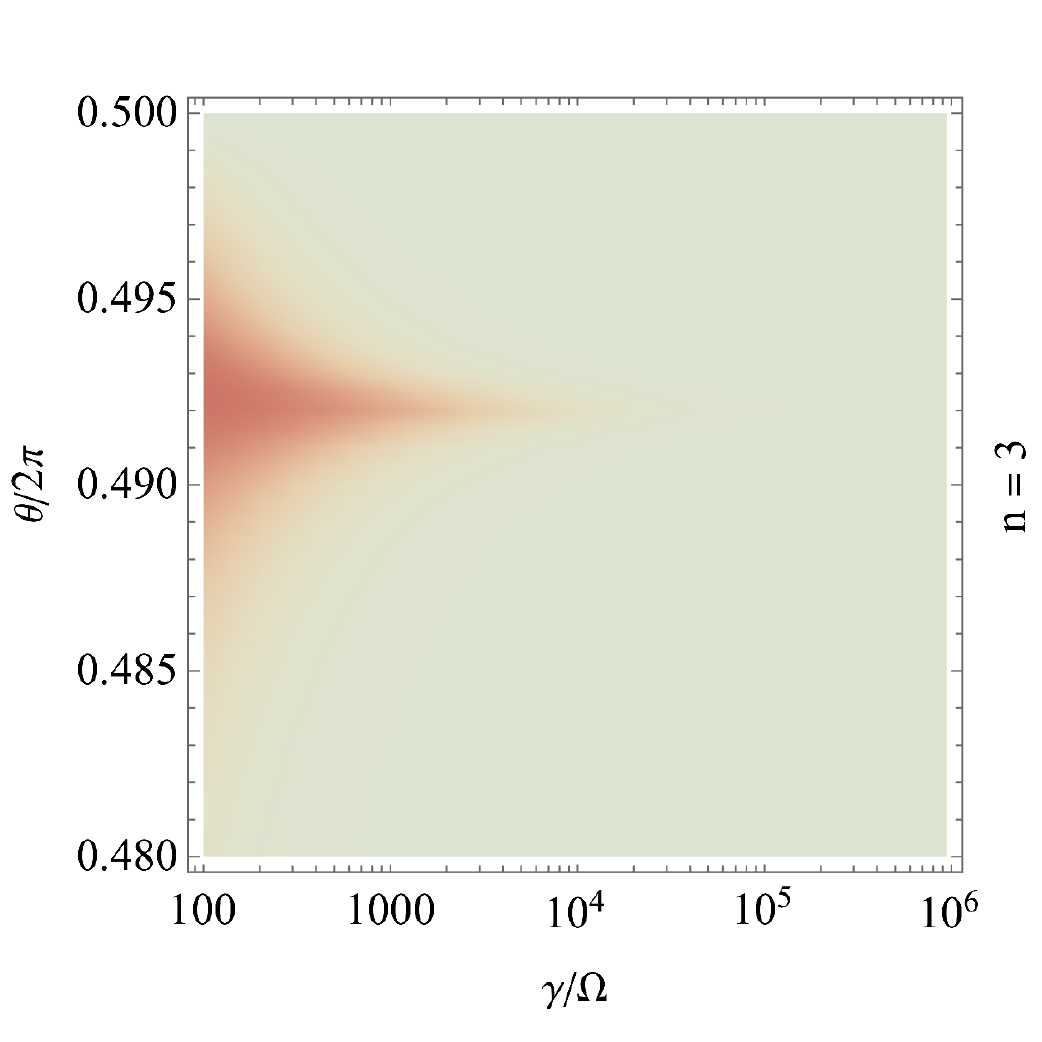}\\
        \includegraphics[width=0.23\textwidth]{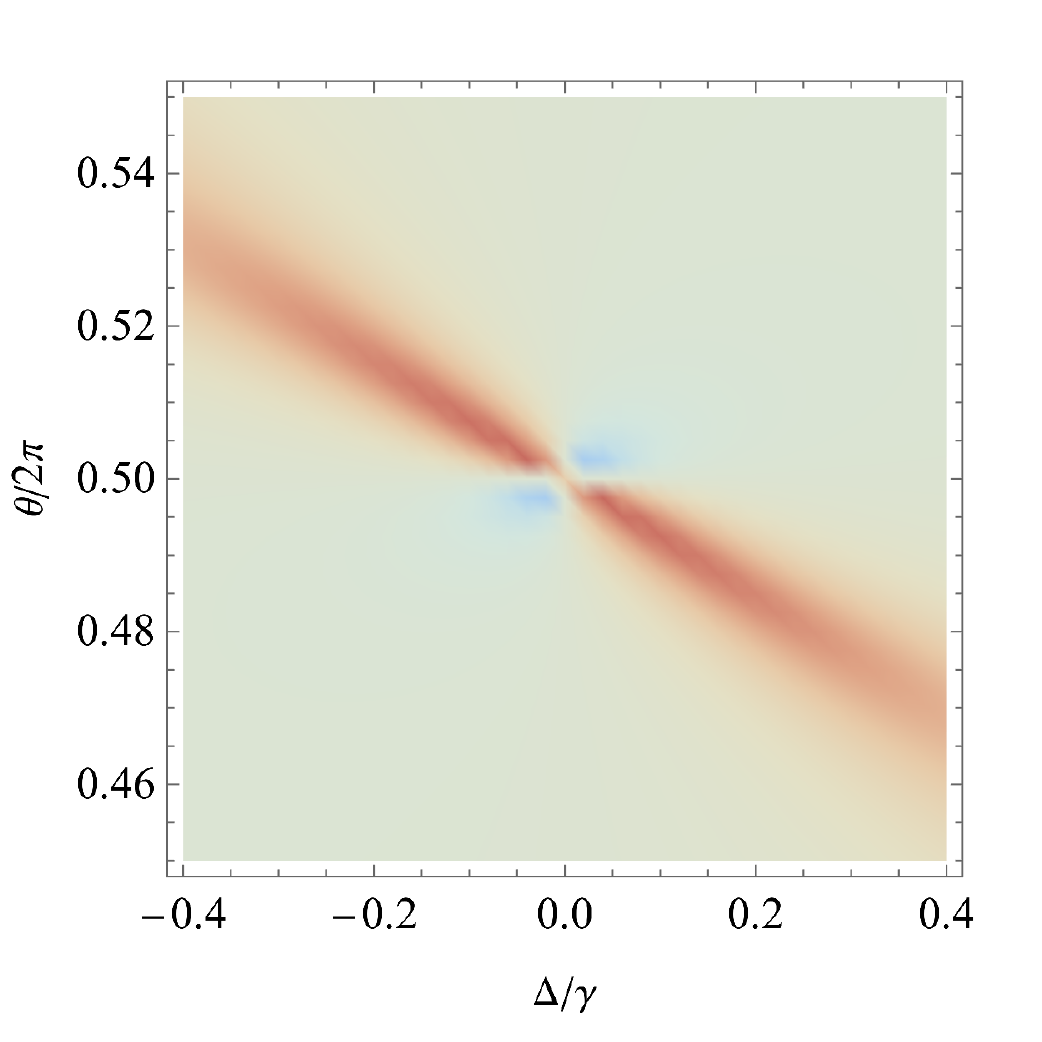}
        \includegraphics[width=0.23\textwidth]{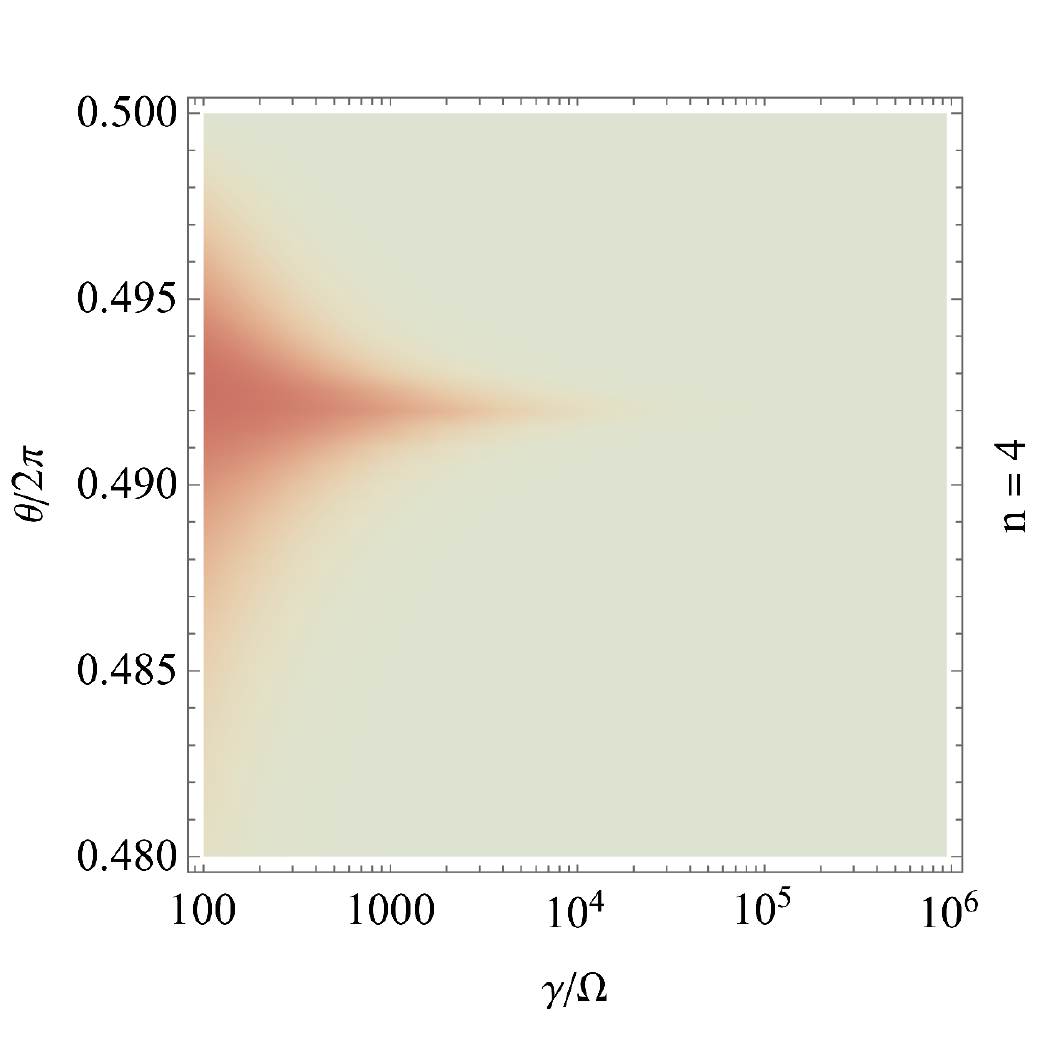}\\
        \includegraphics[width=0.23\textwidth]{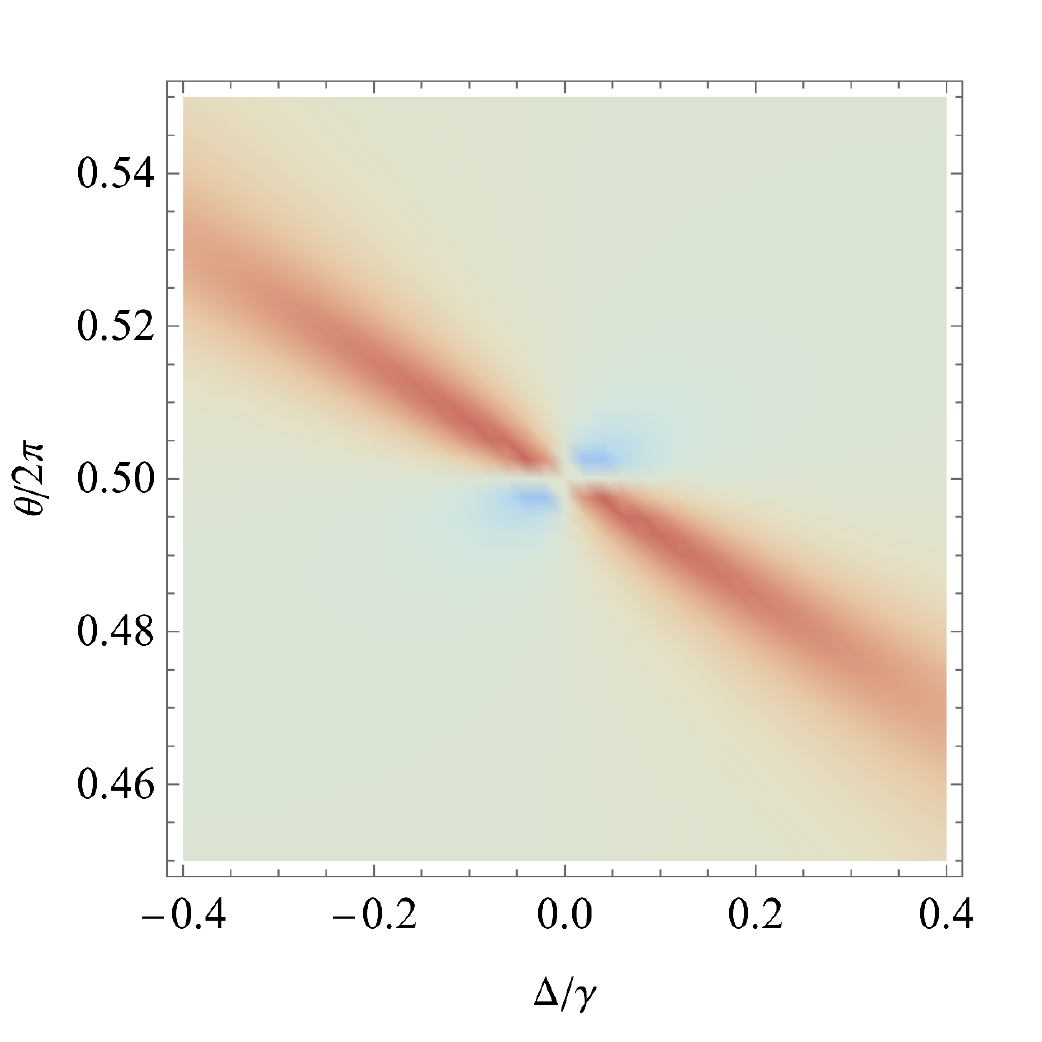}
        \includegraphics[width=0.23\textwidth]{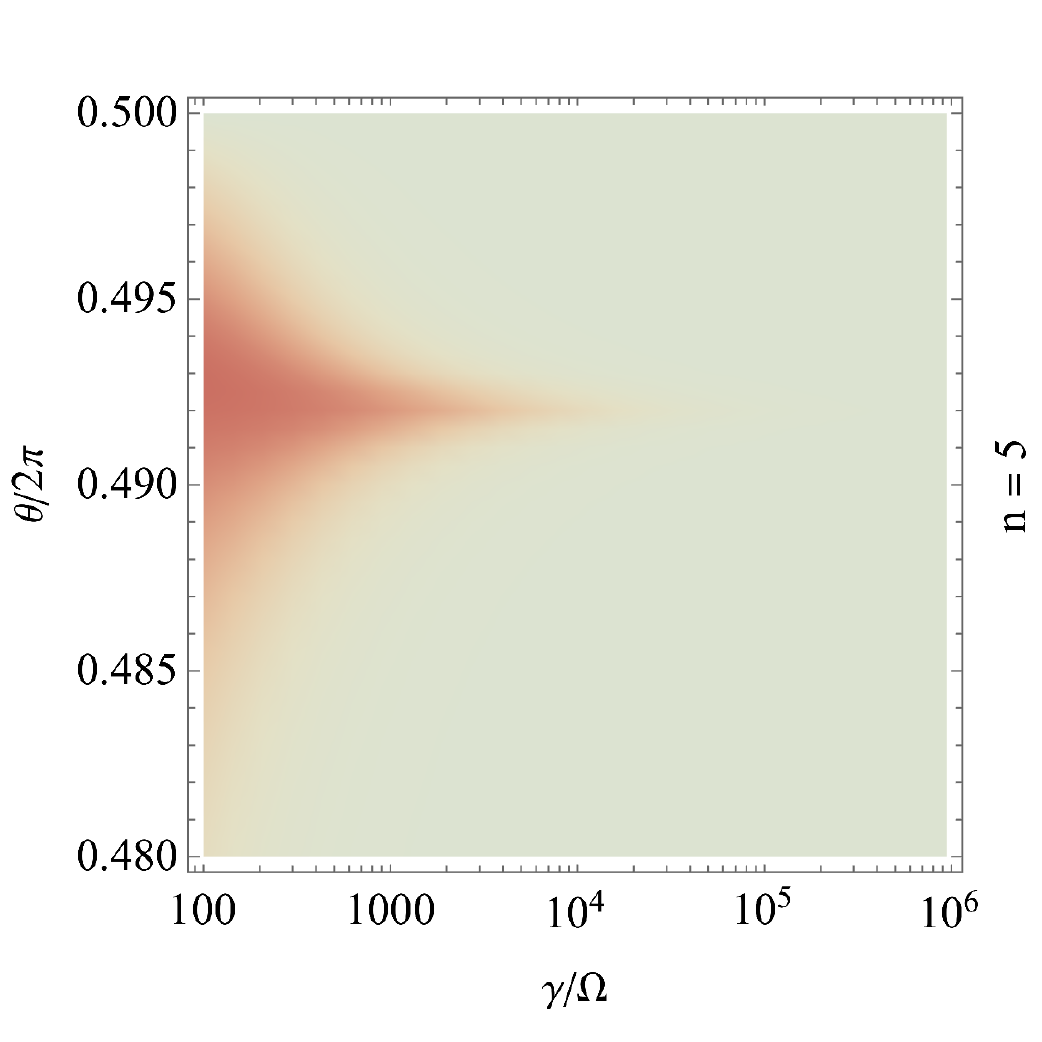}\\
        \includegraphics[width=0.23\textwidth]{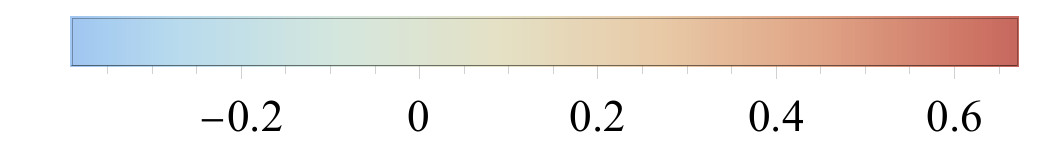}
    \caption{The diode rectification $\mathcal{R}$ for different Fock states ($n=1,2,3,4$ and 5 photons, respectively from top to bottom). Left panels: the rectification factors are computed as a function of the detuning $\Delta = \omega_0 - \omega_1$, and the phase $\theta = \omega_0 d/v_g$ while the bandwidth of the input pulse is fixed, ($\gamma/\Omega = 10^{2}$. On the right panels, we plot the rectification factors for a fixed detuning, $\Delta/\gamma = 0.1$ while we vary the bandwidth $\Omega$.}
    \label{fig:eff}
    \vspace{-2cm}
\end{figure}

\begin{figure}
    \centering
        \includegraphics[width=0.48\textwidth]{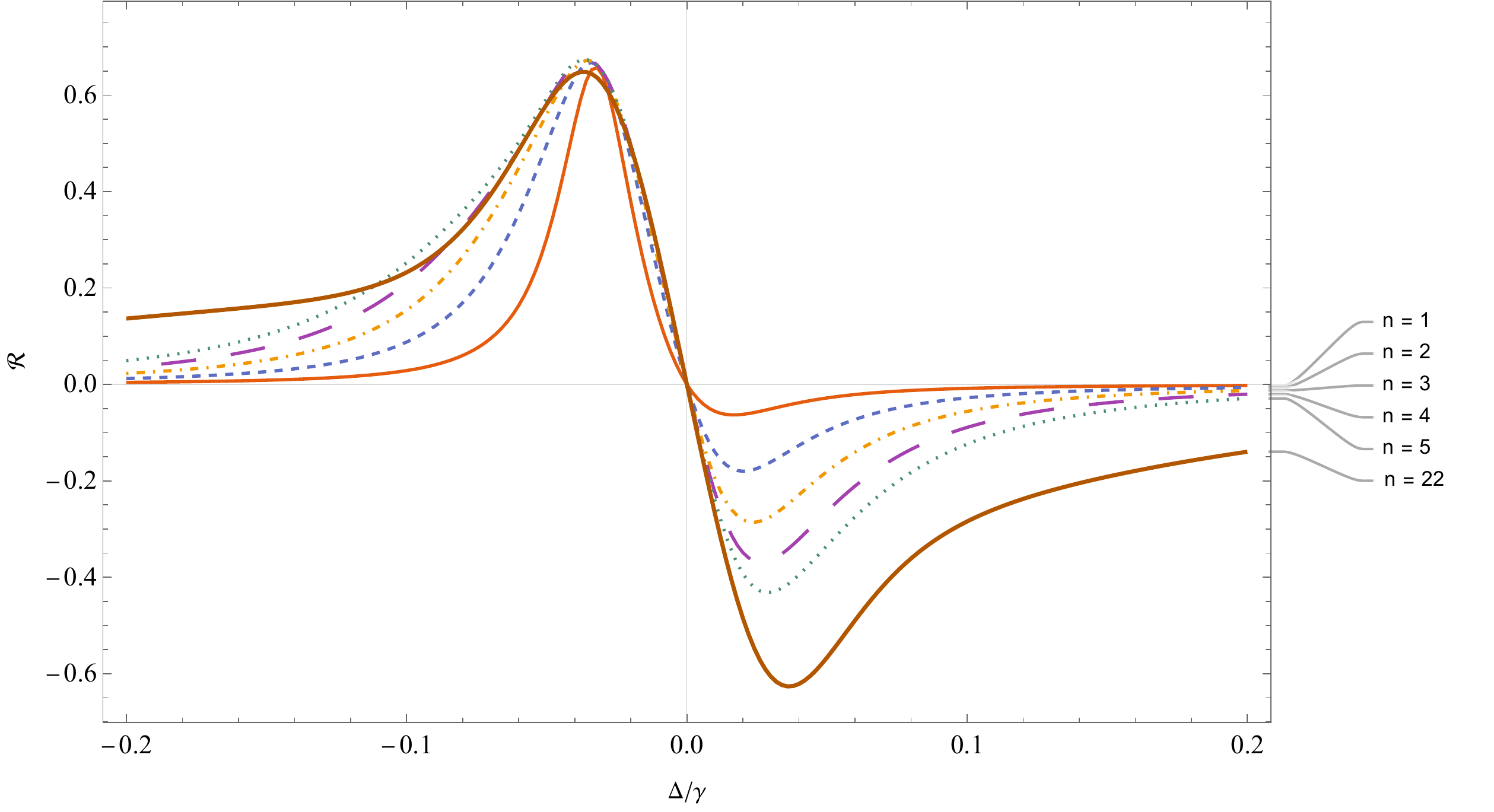}
        \vspace{-2mm}
    \caption{The diode rectification $\mathcal{R}$ for different Fock states ($n=1,2,3,4, 5$ and 22 photons). The pulse length is fixed ($\Omega/\gamma = 10^{-2}$).  The rectification factors are computed as a function of the detuning $\Delta = \omega_0 - \omega_1$. For the case of $n=1$, the phase $\theta$ is optimized in order to have the strongest negativity of the rectification factor. Ergo, we set  $\theta / 2\pi = 0.5025$.  Then we used the same value for the other cases. Here, we show the case of $n=22$ because for this Fock state we observed the strongest negative rectification.}
    \label{fig:negeff}
\end{figure}

\section{Beyond the monochromatic case}

In this section, we consider the case of a pulse with a finite length. However, for our approximation \eqref{ss:assume} to be valid, we need the pulse to be long enough such that the transient regime remains negligible. We shall set the bandwidth to be two orders of magnitude smaller than the decay rate to the waveguide, $\Omega/\gamma = 10^{-2}$. For such a value, the necessary average number of photons to be in the middle of the plateau for the coherent state should be $\bar{n}_{\textrm{opt}} \approx 2$.

\subsection{The device is not sensitive to coherences between Fock states}

We first prove that coherences between Fock states don't play any role in the dynamics of the rectification device. Consider as input state a generic superposition
\ba
\ket{\psi}&=&\sum_{n=0}^\infty c_n\ket{n} \;,
\ea 
with $\ket{n}$ standing for the Fock state with $n$ photons. For definiteness, we look at a special case, but the argument is the same for any of the calculations of interest here.

We assume therefore that the state $\ket{\psi}$ is prepared in mode $\hat{A}$, i.e.~the input state is $\ket{\psi_a}$ with the Fock states given by \cite{Alex90s,Alex90ss}
\begin{equation}\label{state:Fockn}
    \ket{n} = \frac{(\hat{A}^{\dagger})^n}{\sqrt{n!}} \ket{\text{\o{}}}\,.
\end{equation} Then we look at the number of reflected photons
\begin{equation}\label{eq:NnmAll}
    N_{b}^{\psi} (t) = 
    \sum_{m,n}c_nc_m^*\int_{0}^{\infty}\dd\omega\bra{m_a}\hat{b}^{\dagger}_{\omega}(t)\hat{b}_{\omega}(t)\ket{n_a}\,.
\end{equation}
Since we are working within the rotating-wave approximation [Eq.~\eqref{Hdip}], the number of excitations is conserved at all times; and the operator $\hat{b}^{\dagger}_{\omega}(t)\hat{b}_{\omega}(t)$ counts the excitations in part of the system (which may have a photonic and an atomic component). If $m\neq n$, either the number of excitations in that part, or the number of excitations in the rest, will be different between $\ket{n_a}$ and $\ket{m_a}$. Thus, $\bra{m_a}\hat{b}^{\dagger}_{\omega}(t)\hat{b}_{\omega}(t)\ket{n_a}=\bra{n_a}\hat{b}^{\dagger}_{\omega}(t)\hat{b}_{\omega}(t)\ket{n_a}\delta_{m,n}$.

Thus, coherences across the Fock basis in the input state do not play any role. The rectifying device is fully characterised by studying its behavior on each Fock state $\ket{n}$ separately, as we are going to do next.

\subsection{Photon-number pulses}

Let us now study rectification for Fock states \eqref{state:Fockn}. The corresponding flux for our chosen pulse \eqref{xitau} can be computed from $[\hat{a}_\tau(0),(A^{\dagger})^n] = n \xi (\tau) (A^{\dagger})^{n-1}$, which implies $\hat{a}_\tau(0)\ket{n} = \sqrt{n} \xi(\tau) \ket{n-1}$, and finally \ba
F&=&\expval{\hat{a}_\tau^\dagger(0)\hat{a}_\tau(0)}{n} = \frac{n\Omega}{2}\,.
\ea

In the Hamiltonian \eqref{Hdip}, the operator $\hat{b}_{\omega}$ evolves as
\begin{equation}\label{btomega}
\begin{split}
    \hat{b}_{\omega} (t) =& \;\hat{b}_{\omega} (0) + g_{\omega}^{(1)} \int_{0}^{t} \dd t' \hat{\sigma}_{-}^{(1)}(t') e^{i(\omega-\omega_1)t'} \\ 
    & + g_{\omega}^{(2)} \int_{0}^{t} \dd t' \hat{\sigma}_{-}^{(2)}(t') e^{i[(\omega-\omega_2)t' + \omega d / v_g]}\;.
\end{split}
\end{equation} As an example of what we have to compute, the number of reflected photons when the light comes in from the left is
\begin{equation}\label{Nb:coh}
\begin{split}
    N_{\text{ref}}^{(n)} (t) =& \int_{0}^{\infty}\dd\omega\bra{n_a}\hat{b}^{\dagger}_{\omega} (t)\hat{b}_{\omega} (t)\ket{n_a}\;,\\
    =&  \frac{\gamma}{2} \int_{0}^{t}\dd t' \Big\{ \expval{\hat{\sigma}_z^{(1)}(t')}  +  \expval{\hat{\sigma}_z^{(2)}(t')} + 2   \\
    & + 2  \left[ \expval{\hat{\sigma}_{-}^{(1)}(t')\hat{\sigma}_{+}^{(2)}(t')} e^{-i (\Delta_{12} t' - \omega_2 \mu)} + c.c.\right] 
    \Big\} \;,
\end{split}
\end{equation}
where $\Delta_{12} =\omega_1 - \omega_2$ and $\mu =  d / v_g$.

The dynamics is given by the closed set of Heisenberg equations of motion presented in \cite{Alex}, which we reproduce for completeness in Appendix \ref{app:heis}. We solve them numerically in the steady state regime, for various input states, to obtain the desired expectation values \eqref{ss:assume}.

In Fig.~\ref{fig:eff}, the rectification factor for various input states is illustrated. In the case of a fixed bandwidth, $\gamma/\Omega = 10^{2} $, we can see that we have a significant rectification for all the Fock states. For all these cases we observe a maximum rectification near 66\%. However, the range of inter-atomic distances, $d$, and detunings, $\Delta$, for which the strongest directionalities are observed, is narrower for Fock states with smaller number of photons. On the other hand, when we decrease the bandwidth, the directionality is lost for all the values of the distance between the qubits, $d$. For the case of a single photon, we observe that the rectification fades for a bandwidth between $\gamma/\Omega \sim 10^{3}$ and $10^{4}$. For an increasing number of photons, the rectification appears to be more robust against the narrowing of the bandwidth. For instance, in the case of 5 photons, the rectification of the device vanishes between $\gamma/\Omega \sim 10^{4}$ and $10^{5}$. Also, a similar behaviour can be observed for a fixed inter-atomic distance while varying the detuning. This is in agreement with the result in \cite{Alex} that the device becomes symmetric for single photon inputs within the monochromatic limit.

Interestingly, we observe that the rectification factor takes negative values (e.g. near $\theta/2\pi = 0.5025$ and $\Delta/\gamma = 0.04$). According to the definition \eqref{DiodeEff}, this means that the direction of the diode became reversed. In Fig.~\ref{fig:negeff}, we can see the maximum positive value of the rectification factor, for all the considered number states, are between 0.6 and 0.66. However, in the negative regime, the maximum efficiency of the device seems to depend on the number of photons. The strongest negativity is observed for the case of 22 photons. In other words, increasing or decreasing the number of photons results in weaker negative rectifications. Note that in this optimal case, $n=22$, the behaviour of the device is anti-symmetric. Hence, by changing the detuning between positive and negative values, one can flip the diode's preferred direction.

\section{Conclusions}

In this paper, we have clarified (and occasionally rectified, pun intended) the rectifying functionality of the simple optical diode sketched in Fig.~\ref{fig:diode}. We have found that its behavior is completely determined by the behavior on the Fock states, and that some single-photon rectification does happen when the pulses are of finite length.

This study was done under two main assumptions: (i) that the pulses are monomode and long compared to decay time of atomic excitations, and (ii) that the rotating-wave approximation holds. To go beyond the first and consider composite and/or short pulses, one would have to solve the time-dependent Heisenberg equations of motion. Removing the second assumption may be of interest too, given that the diode is naturally implemented with superconducting qubits \cite{Clemens}, a platform in which ultrastrong coupling can be reached \cite{Zueco,Niemczyk}.

\section*{Acknowledgment}
This research is supported by the National Research Foundation and the Ministry of Education, Singapore, under the Research Centres of Excellence programme. Part of this work was done when S.L.~was on exchange at the National University of Singapore in a NGNE programme. S.L.~acknowledges financial support from the School of Physics and the Qian Xuesen Honors College, Xi'an Jiaotong University.

\appendix

\section{Comparative study of different definitions of rectification factors}
\label{app:defs}

\begin{figure}[t]
    \centering
        \includegraphics[width=0.23\textwidth]{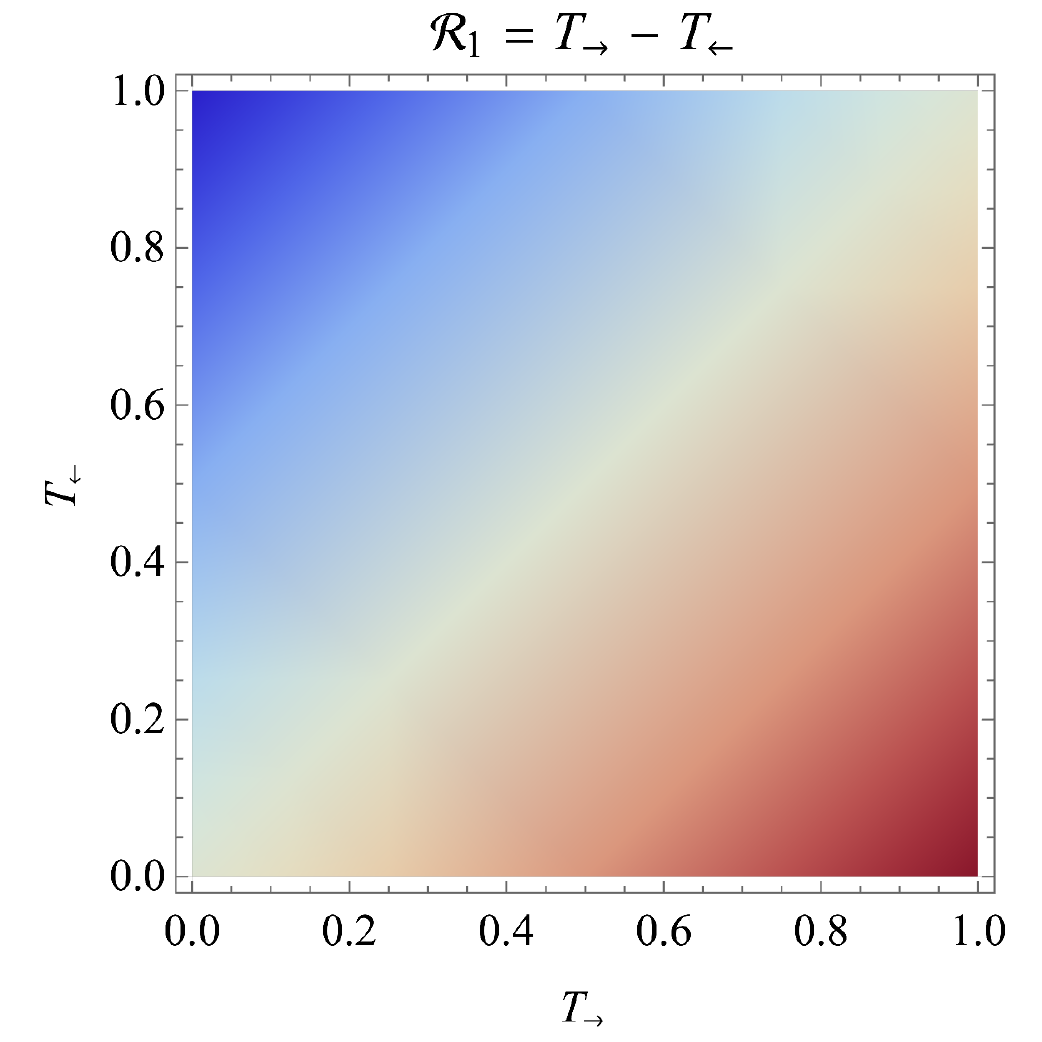}
        \includegraphics[width=0.23\textwidth]{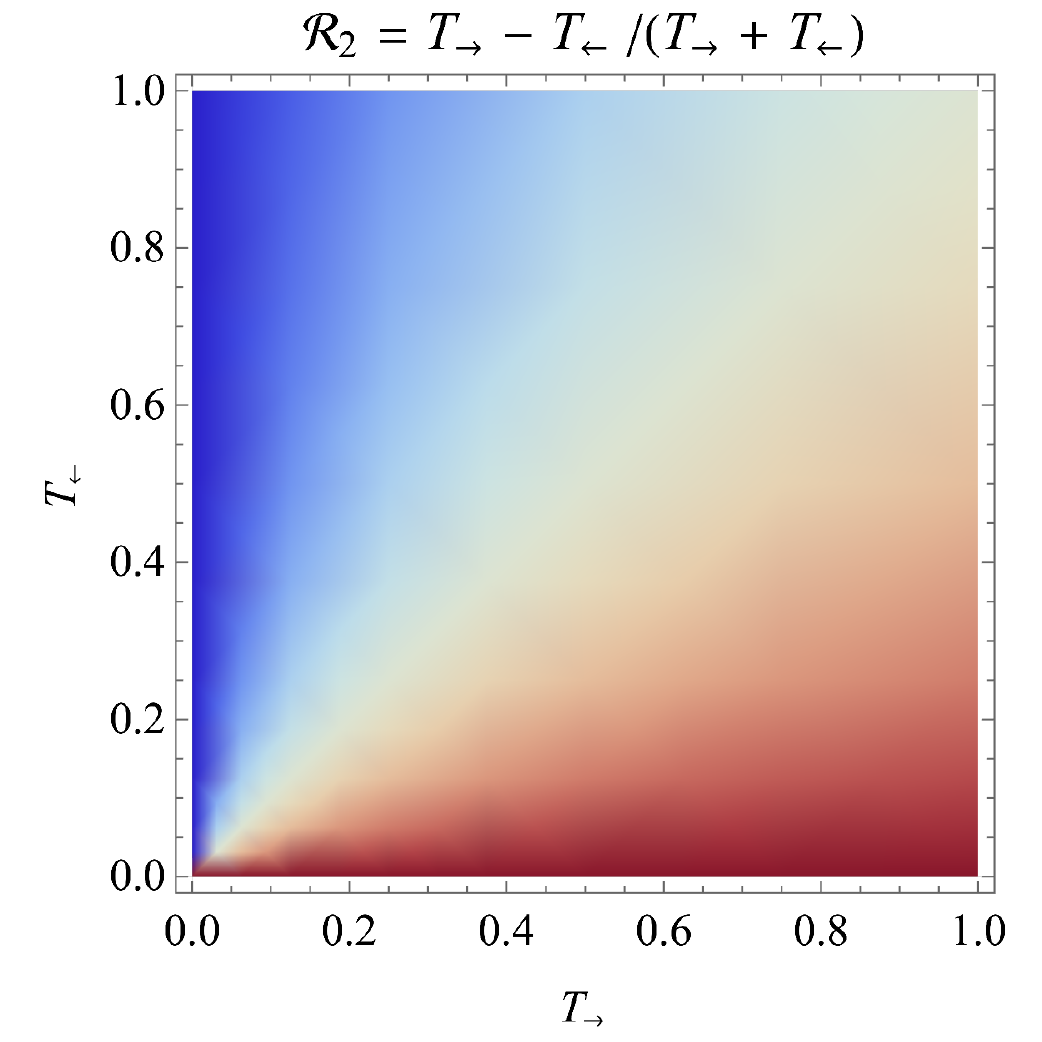}\\
        \includegraphics[width=0.23\textwidth]{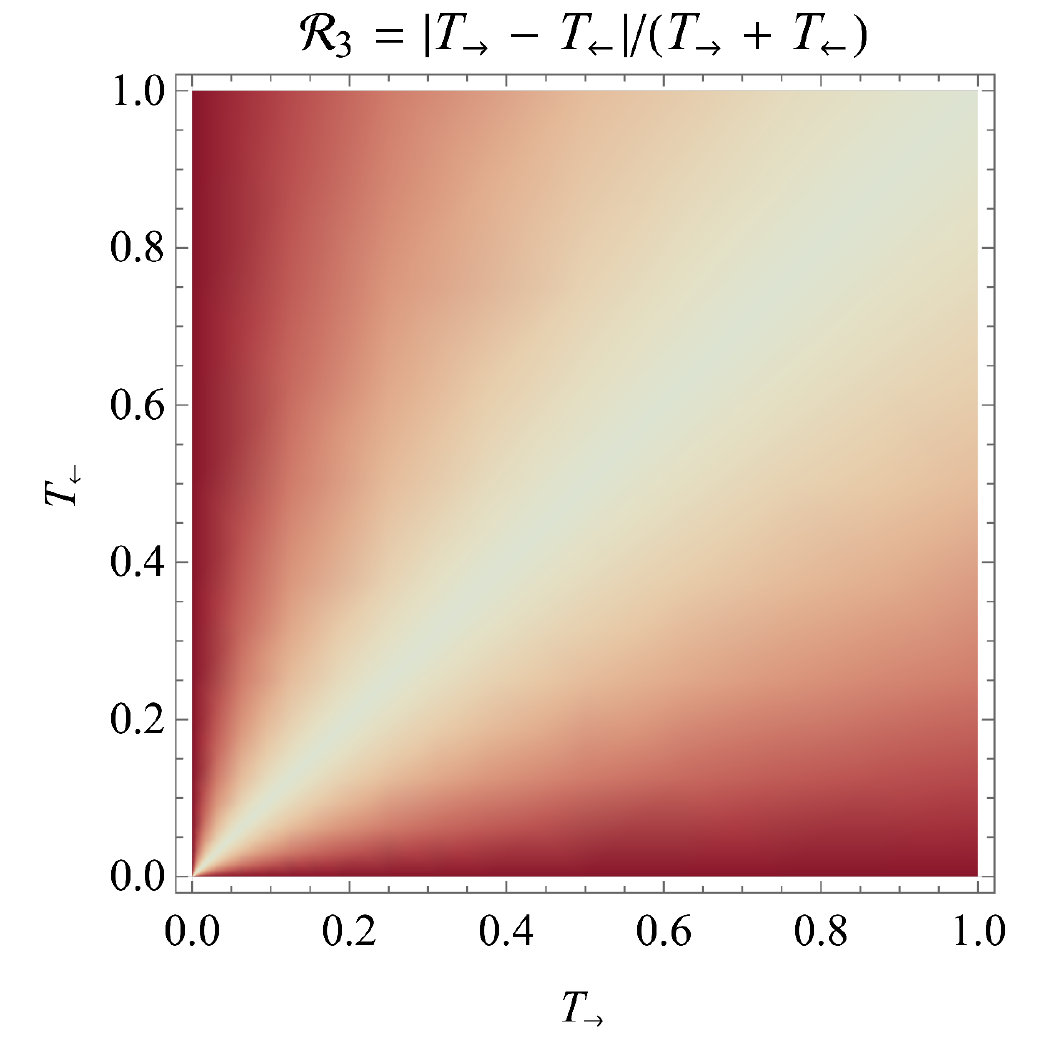}
        \includegraphics[width=0.23\textwidth]{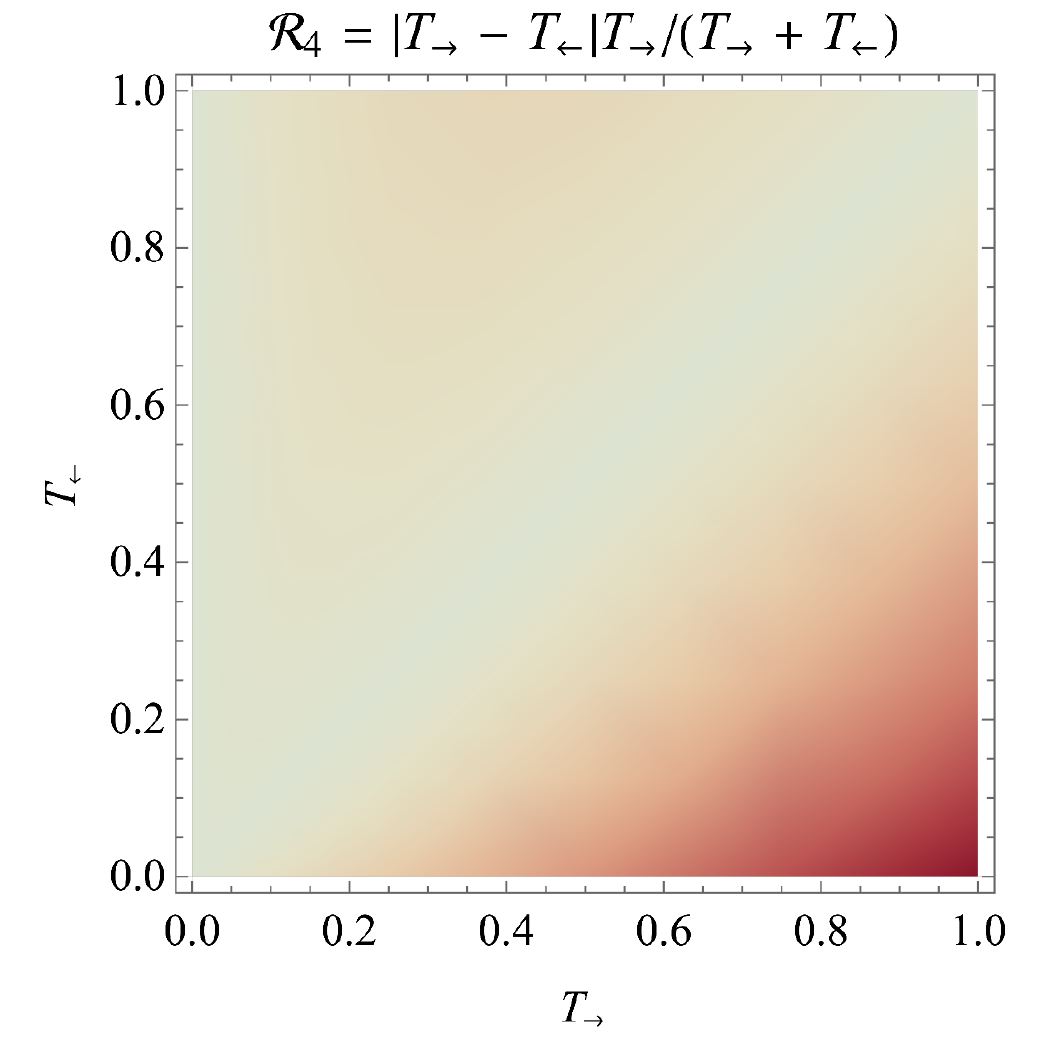}\\
        \vspace{-2mm}
        \includegraphics[width=0.23\textwidth]{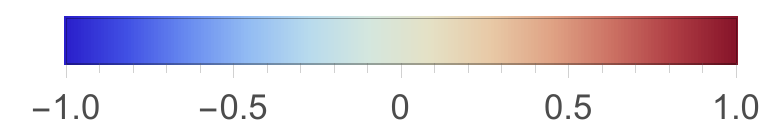}
    \caption{The diode rectification factor $\mathcal{R}$ using different definitions.}
    \label{fig:effdef}
\end{figure}

\begin{figure}[t]
    \centering
        \includegraphics[width=0.48\textwidth]{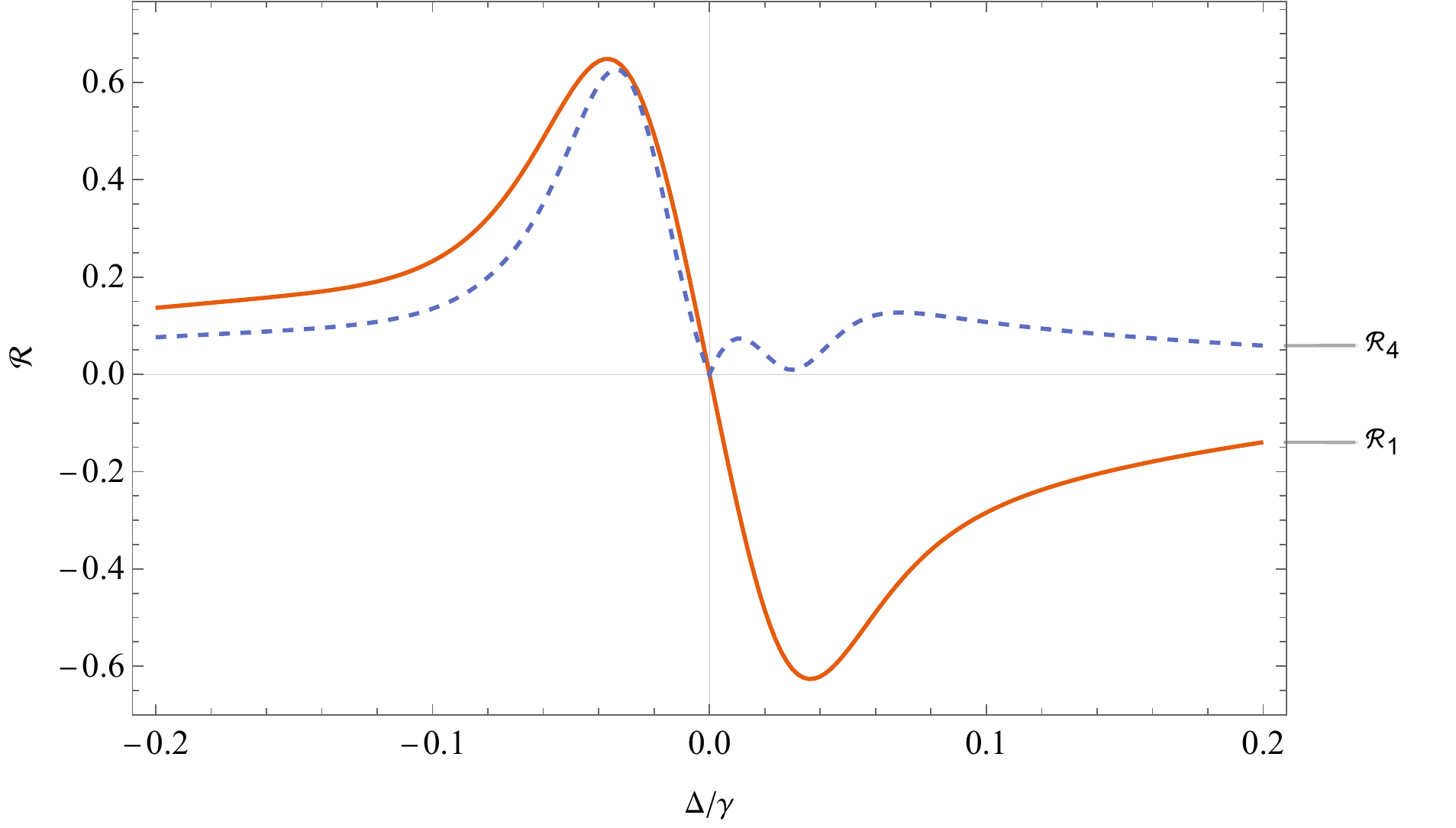}
    \caption{The diode rectification factor for $n = 22$. We plot the rectification as a function of the detuning $\Delta$ while fixing the phase $\theta / 2\pi = 0.5025$. The red solid line correspond the definition $\mathcal{R}_1$ while the blue dashed line corresponds to $\mathcal{R}_4$.}
    \label{fig:effneg}
\end{figure}

For the device we are considering in this work, there are a few possible definitions for quantifying its directionality. In this work, we have adopted Eq.~\eqref{DiodeEff}:
\begin{equation}\label{R1}
    \mathcal{R}_1 = T_{\rightarrow} - T_{\leftarrow}\;.
\end{equation}
It can take values between $-1$ and $+1$. A rectification factor equal to zero corresponds to a completely symmetric device, while $\pm 1$ stands for the perfect rectifying device. Besides its simplicity, $\mathcal{R}_1$ has the nice symmetry $\mathcal{R}_1 = R_{\leftarrow}-R_{\rightarrow}$ with $R_{.}=1-T_{.}$. In words, it is the only definition listed that yields the same number for the equivalent descriptions ``transmit more an input from the left than one from the right'' and ``reflect more an input from the right than one from the left''. 

In the literature, one can find other definitions: for instance,
\begin{equation}\label{R2}
    \mathcal{R}_2 = \frac{T_{\rightarrow} - T_{\leftarrow}}{T_{\rightarrow} + T_{\leftarrow}}\;,
\end{equation}
\begin{equation}\label{R3}
    \mathcal{R}_3 = \frac{\abs{T_{\rightarrow} - T_{\leftarrow}}}{T_{\rightarrow} + T_{\leftarrow}}\;,
\end{equation}
\begin{equation}\label{R4}
    \mathcal{R}_4 = \frac{\abs{T_{\rightarrow} - T_{\leftarrow}}}{T_{\rightarrow} + T_{\leftarrow}}T_{\rightarrow}\;.
\end{equation}
The definition in Eq.~\eqref{R2}, has been first proposed in \cite{Lepri}. In \cite{auffeves2014}, the definitions \eqref{R3} and \eqref{R4} were used while referring to $\mathcal{R}_3$ as a rectification factor and to $\mathcal{R}_4$ as the efficiency. In other works \cite{Alex,Fang,Auffeves2016,Vienna2016,Ordonez,Combes,Clemens,Eduardo} different definitions were used, including but not limited to (\ref{R2}-\ref{R4})

Interestingly, all definitions (\ref{R1}-\ref{R4}) show different behaviours. In Fig.~\ref{fig:effdef} we can see that both $\mathcal{R}_1$ and $\mathcal{R}_2$ are anti-symmetric, while $\mathcal{R}_3$ is symmetric and $\mathcal{R}_4$ is asymmetric with respect to the diagonal ($T_{\rightarrow} = T_{\leftarrow}$).

The main differences between all these definitions are as follows: when we do not divide by $T_{\rightarrow} + T_{\leftarrow}$ the rectification factor indicates a perfect behaviour, $\mathcal{R}_1 = 1$, if and only if $T_{\rightarrow} = 1$ and $T_{\rightarrow} = 0$. On the other hand, if we divide by $T_{\rightarrow} + T_{\leftarrow}$ as in the remaining 3 definitions, the values become somewhat inflated. In this case, the perfect rectification, $\mathcal{R}_j = 1$ while $j \in \{2,3,4\}$, may correspond to $T_{\leftarrow} = 0$ and $T_{\rightarrow} > 0$. Note that the values inside their respective plot frames in Fig.~\ref{fig:effdef} are also increased. Using $\mathcal{R}_1$ and $\mathcal{R}_2$, the sign the rectification factor give us information about the direction of the diode. On the other hand, $\mathcal{R}_3$ indicates the strength of the rectification but it does not give any information about the preferred direction. The definition $\mathcal{R}_4$, captures the rectification only into one direction. As illustrated in Fig.~\ref{fig:negeff}, we plot the rectification factor using the definitions $\mathcal{R}_1$ and $\mathcal{R}_4$. In this example, we can see that the diode exhibits an anti-symmetric behaviour which is captured only by $\mathcal{R}_1$.

\section{Heisenberg equations of motion}\label{app:heis}

In this appendix, for completeness we reproduce from \cite{Alex} the closed set of Heisenberg equations of motions that describe the dynamics of our model. 
\begin{widetext}
\begin{equation}
\begin{split}
    \frac{d}{dt} \hat{\sigma}_{-}^{(1)} =& - \gamma (\hat{\sigma}_{-}^{(1)} - \hat{\sigma}_{z}^{(1)}\hat{\sigma}_{-}^{(2)}e^{-i (\Delta t - \omega_0 \mu)}) + \sqrt{\gamma} \hat{\sigma}_{z}^{(1)} (\hat{a}_{t}^{} + \hat{b}_{t}^{}) \;,
\end{split}
\end{equation}

\begin{equation}
\frac{d}{dt} \hat{\sigma}_{z}^{(1)} = -2\gamma \left( \mathbb{1} + \hat{\sigma}_{z}^{(1)}\right) - 2 \gamma \left(\hat{\sigma}_{+}^{(1)}\hat{\sigma}_{-}^{(2)}e^{-i (\Delta t - \omega_0 \mu)} + \textbf{H.c.}\right) - 2 \sqrt{\gamma} \left[ \hat{\sigma}_{+}^{(1)} (\hat{a}_{t}^{} + \hat{b}_{t}^{}) + \textbf{H.c.}\right] \;,
\end{equation}
\begin{equation}
\frac{d}{dt} \hat{\sigma}_{-}^{(2)} =  - \gamma \left(\hat{\sigma}_{-}^{(2)} - \hat{\sigma}_{z}^{(2)}\hat{\sigma}_{-}^{(1)}e^{i (\Delta t + \omega_0 \mu)}\right) + 
     \sqrt{\gamma} \hat{\sigma}_{z}^{(2)} \left(\hat{a}_{t-\mu}^{}e^{i \omega_1 \mu} + \hat{b}_{t+\mu}^{}e^{-i \omega_1 \mu}\right)e^{i \Delta t} \;,
\end{equation}
\begin{equation}
\frac{d}{dt} \hat{\sigma}_{z}^{(2)} = -2\gamma \left( \mathbb{1} + \hat{\sigma}_{z}^{(2)}\right) - 2 \gamma \left(\hat{\sigma}_{+}^{(2)}\hat{\sigma}_{-}^{(1)}e^{i (\Delta t + \omega_1 \mu)} + \textbf{H.c.}\right) - 2 \sqrt{\gamma} \left[ \hat{\sigma}_{+}^{(2)} (\hat{a}_{t-\mu}^{}e^{i \omega_1 \mu} + \hat{b}_{t+\mu}^{}e^{-i \omega_1 \mu})e^{i \Delta t} + \textbf{H.c.}\right] \;,
\end{equation}
\begin{equation}
\begin{split}
\frac{d}{dt} (\hat{\sigma}_{z}^{(1)}\hat{\sigma}_{-}^{(2)}) =& -3\gamma \hat{\sigma}_{z}^{(1)}\hat{\sigma}_{-}^{(2)} - 2\gamma \hat{\sigma}_{-}^{(2)} - \gamma \hat{\sigma}_{-}^{(1)}e^{i (\Delta t - \omega_0 \mu)} - \gamma \hat{\sigma}_{-}^{(1)}\hat{\sigma}_{z}^{(2)}e^{i \Delta t} \left( e^{-i \omega_0 \mu} +e^{ i \omega_1 \mu} \right) \\
&- 2 \sqrt{\gamma}\left[ \hat{\sigma}_{+}^{(1)}\hat{\sigma}_{-}^{(2)}(\hat{a}_{t}^{} + \hat{b}_{t}^{}) + (\hat{a}_{t}^{\dagger} + \hat{b}_{t}^{\dagger})\hat{\sigma}_{-}^{(1)}\hat{\sigma}_{-}^{(2)}\right] + \sqrt{\gamma} \hat{\sigma}_{z}^{(1)}\hat{\sigma}_{z}^{(2)} e^{i \Delta t}\left(\hat{a}_{t-\mu}^{}e^{i \omega_1 \mu} + \hat{b}_{t+\mu}^{}e^{-i \omega_1 \mu}\right) \;,
\end{split}
\end{equation}
\begin{equation}
\begin{split}
\frac{d}{dt} (\hat{\sigma}_{-}^{(1)}\hat{\sigma}_{z}^{(2)}) =& -3\gamma \hat{\sigma}_{-}^{(1)}\hat{\sigma}_{z}^{(2)} - 2\gamma \hat{\sigma}_{-}^{(1)} - \gamma \hat{\sigma}_{-}^{(2)}e^{-i (\Delta t + \omega_0 \mu)} - \gamma \hat{\sigma}_{z}^{(1)}\hat{\sigma}_{-}^{(2)}e^{-i \Delta t} \left( e^{i \omega_0 \mu} +e^{- i \omega_1 \mu} \right) \\
&- 2 \sqrt{\gamma}\left[ \hat{\sigma}_{-}^{(1)}\hat{\sigma}_{+}^{(2)}
\left(\hat{a}_{t-\mu}^{}e^{i \omega_1 \mu} + \hat{b}_{t+\mu}^{}e^{-i \omega_1 \mu}\right)e^{i \Delta t} 
+ \left( \hat{a}_{t-\mu}^{\dagger}e^{-i \omega_1 \mu} +\hat{b}_{t+\mu}^{\dagger}e^{i \omega_1 \mu} \right)e^{-i \Delta t}  \hat{\sigma}_{-}^{(1)}\hat{\sigma}_{-}^{(2)}\right] \\
&+ \sqrt{\gamma} \hat{\sigma}_{z}^{(1)}\hat{\sigma}_{z}^{(2)} 
\left(\hat{a}_{t}^{} + \hat{b}_{t}^{}\right) \;,
\end{split}
\end{equation}
\begin{equation}
\begin{split}
\frac{d}{dt} (\hat{\sigma}_{+}^{(1)}\hat{\sigma}_{-}^{(2)}) =& 
- 2 \gamma \hat{\sigma}_{+}^{(1)}\hat{\sigma}_{-}^{(2)} + \frac{1}{2} \gamma e^{i \Delta t}
\left( \hat{\sigma}_{z}^{(1)}e^{-i \omega_0 \mu} + \hat{\sigma}_{z}^{(2)}e^{i \omega_0 \mu} \right) 
+ \frac{1}{2} \gamma \hat{\sigma}_{z}^{(1)}\hat{\sigma}_{z}^{(2)} e^{i \Delta t} 
\left( e^{i \omega_0 \mu} +e^{ -i \omega_1 \mu} \right) \\
& + \sqrt{\gamma} \left(\hat{a}_{t}^{\dagger} + \hat{b}_{t}^{\dagger}\right)\hat{\sigma}_{z}^{(1)}\hat{\sigma}_{-}^{(2)}
+ \sqrt{\gamma} \hat{\sigma}_{+}^{(1)}\hat{\sigma}_{z}^{(2)}e^{i \Delta t} \left(\hat{a}_{t-\mu}^{}e^{i \omega_1 \mu} + \hat{b}_{t+\mu}^{}e^{-i \omega_1 \mu}\right) \;,
\end{split}
\end{equation}
\begin{equation}
\frac{d}{dt} (\hat{\sigma}_{-}^{(1)}\hat{\sigma}_{-}^{(2)}) = -2 \gamma \hat{\sigma}_{-}^{(1)}\hat{\sigma}_{-}^{(2)} + \sqrt{\gamma} \hat{\sigma}_{z}^{(1)}\hat{\sigma}_{-}^{(2)}\left(\hat{a}_{t}^{} + \hat{b}_{t}^{}\right)
+\sqrt{\gamma} \hat{\sigma}_{-}^{(1)}\hat{\sigma}_{z}^{(2)} \left(\hat{a}_{t-\mu}^{}e^{i \omega_1 \mu} + \hat{b}_{t+\mu}^{}e^{-i \omega_1 \mu}\right) e^{i \Delta t}\;,
\end{equation}
\begin{equation}
\begin{split}
\frac{d}{dt} (\hat{\sigma}_{z}^{(1)}\hat{\sigma}_{z}^{(2)}) =&  -3 \gamma \hat{\sigma}_{z}^{(1)}\hat{\sigma}_{z}^{(2)} - 2 \gamma \left( \hat{\sigma}_{z}^{(1)}+\hat{\sigma}_{z}^{(2)} \right) + 2 \gamma \left[ \hat{\sigma}_{+}^{(1)}\hat{\sigma}_{-}^{(2)} e^{-i \Delta t}\left( e^{i \omega_0 \mu} +e^{ -i \omega_1 \mu} \right) + \textbf{H.c.} \right] \\
& - 2\sqrt{\gamma} \left[ \hat{\sigma}_{+}^{(1)}\hat{\sigma}_{z}^{(2)}\left(\hat{a}_{t}^{} + \hat{b}_{t}^{}\right) +  \textbf{H.c.} \right] - 2 \sqrt{\gamma} \left[ \hat{\sigma}_{z}^{(1)}\hat{\sigma}_{+}^{(2)} \left(\hat{a}_{t-\mu}^{}e^{i \omega_1 \mu} + \hat{b}_{t+\mu}^{}e^{-i \omega_1 \mu}\right) e^{i \Delta t} + \textbf{H.c.}\right] \;.
\end{split}
\end{equation}
\end{widetext}

\bibliography{refs}

\begin{thebibliography}{22}%
\makeatletter
\providecommand \@ifxundefined [1]{%
 \@ifx{#1\undefined}
}%
\providecommand \@ifnum [1]{%
 \ifnum #1\expandafter \@firstoftwo
 \else \expandafter \@secondoftwo
 \fi
}%
\providecommand \@ifx [1]{%
 \ifx #1\expandafter \@firstoftwo
 \else \expandafter \@secondoftwo
 \fi
}%
\providecommand \natexlab [1]{#1}%
\providecommand \enquote  [1]{``#1''}%
\providecommand \bibnamefont  [1]{#1}%
\providecommand \bibfnamefont [1]{#1}%
\providecommand \citenamefont [1]{#1}%
\providecommand \href@noop [0]{\@secondoftwo}%
\providecommand \href [0]{\begingroup \@sanitize@url \@href}%
\providecommand \@href[1]{\@@startlink{#1}\@@href}%
\providecommand \@@href[1]{\endgroup#1\@@endlink}%
\providecommand \@sanitize@url [0]{\catcode `\\12\catcode `\$12\catcode
  `\&12\catcode `\#12\catcode `\^12\catcode `\_12\catcode `\%12\relax}%
\providecommand \@@startlink[1]{}%
\providecommand \@@endlink[0]{}%
\providecommand \url  [0]{\begingroup\@sanitize@url \@url }%
\providecommand \@url [1]{\endgroup\@href {#1}{\urlprefix }}%
\providecommand \urlprefix  [0]{URL }%
\providecommand \Eprint [0]{\href }%
\providecommand \doibase [0]{https://doi.org/}%
\providecommand \selectlanguage [0]{\@gobble}%
\providecommand \bibinfo  [0]{\@secondoftwo}%
\providecommand \bibfield  [0]{\@secondoftwo}%
\providecommand \translation [1]{[#1]}%
\providecommand \BibitemOpen [0]{}%
\providecommand \bibitemStop [0]{}%
\providecommand \bibitemNoStop [0]{.\EOS\space}%
\providecommand \EOS [0]{\spacefactor3000\relax}%
\providecommand \BibitemShut  [1]{\csname bibitem#1\endcsname}%
\let\auto@bib@innerbib\@empty
\bibitem [{\citenamefont {{W. Asavanant, Yu Shiozawa, S. Yokoyama, B.
  Charoensombutamon, H. Emura, R. N. Alexander, S. Takeda, J. Yoshikawa, N. C.
  Menicucci, H. Yonezawa, A. Furusawa}}(2019)}]{Asavanant}%
  \BibitemOpen
  \bibfield  {author} {\bibinfo {author} {\bibnamefont {{W. Asavanant, Yu
  Shiozawa, S. Yokoyama, B. Charoensombutamon, H. Emura, R. N. Alexander, S.
  Takeda, J. Yoshikawa, N. C. Menicucci, H. Yonezawa, A. Furusawa}}},\
  }\href@noop {} {\bibfield  {journal} {\bibinfo  {journal} {Science}\ }\textbf
  {\bibinfo {volume} {366}},\ \bibinfo {pages} {373–376} (\bibinfo {year}
  {2019})}\BibitemShut {NoStop}%
\bibitem [{\citenamefont {{M.V. Larsen, X. Guo, C.R. Breum, J.S.
  Neergaard-Nielsen, \& U. L. Andersen}}(2019)}]{Larsen}%
  \BibitemOpen
  \bibfield  {author} {\bibinfo {author} {\bibnamefont {{M.V. Larsen, X. Guo,
  C.R. Breum, J.S. Neergaard-Nielsen, \& U. L. Andersen}}},\ }\href@noop {}
  {\bibfield  {journal} {\bibinfo  {journal} {Nature}\ }\textbf {\bibinfo
  {volume} {366}},\ \bibinfo {pages} {369–372} (\bibinfo {year}
  {2019})}\BibitemShut {NoStop}%
\bibitem [{\citenamefont {{J. M. Arrazola, V. Bergholm, K. Bredler, T. R.
  Bromley, et al.}}(2021)}]{Juan}%
  \BibitemOpen
  \bibfield  {author} {\bibinfo {author} {\bibnamefont {{J. M. Arrazola, V.
  Bergholm, K. Bredler, T. R. Bromley, et al.}}},\ }\href@noop {} {\bibfield
  {journal} {\bibinfo  {journal} {Nature}\ }\textbf {\bibinfo {volume} {591}},\
  \bibinfo {pages} {54–60} (\bibinfo {year} {2021})}\BibitemShut {NoStop}%
\bibitem [{\citenamefont {{D.M. Pozar}}(1998)}]{Pozar}%
  \BibitemOpen
  \bibfield  {author} {\bibinfo {author} {\bibnamefont {{D.M. Pozar}}},\
  }\href@noop {} {\bibfield  {journal} {\bibinfo  {journal} {Microwave
  Engineering}\ } (\bibinfo {year} {Wiley, Crawfordsville, 1998})}\BibitemShut
  {NoStop}%
\bibitem [{\citenamefont {{F. Fratini, E. Mascarenhas, L. Safari, J-Ph. Poizat,
  D. Valente, A. Auffèves, D. Gerace, and M. F.
  Santos}}(2014)}]{auffeves2014}%
  \BibitemOpen
  \bibfield  {author} {\bibinfo {author} {\bibnamefont {{F. Fratini, E.
  Mascarenhas, L. Safari, J-Ph. Poizat, D. Valente, A. Auffèves, D. Gerace,
  and M. F. Santos}}},\ }\href@noop {} {\bibfield  {journal} {\bibinfo
  {journal} {Phys. Rev. Lett.}\ }\textbf {\bibinfo {volume} {113}},\ \bibinfo
  {pages} {243601} (\bibinfo {year} {2014})}\BibitemShut {NoStop}%
\bibitem [{\citenamefont {{Jibo Dai, Alexandre Roulet, Huy Nguyen Le, and
  Valerio Scarani}}(2015)}]{Alex}%
  \BibitemOpen
  \bibfield  {author} {\bibinfo {author} {\bibnamefont {{Jibo Dai, Alexandre
  Roulet, Huy Nguyen Le, and Valerio Scarani}}},\ }\href@noop {} {\bibfield
  {journal} {\bibinfo  {journal} {Phys. Rev. A}\ }\textbf {\bibinfo {volume}
  {92}},\ \bibinfo {pages} {063848} (\bibinfo {year} {2015})}\BibitemShut
  {NoStop}%
\bibitem [{\citenamefont {{E. Mascarenhas, M. F. Santos, A. Auffeves, and D.
  Gerace}}(2016)}]{Auffeves2016}%
  \BibitemOpen
  \bibfield  {author} {\bibinfo {author} {\bibnamefont {{E. Mascarenhas, M. F.
  Santos, A. Auffeves, and D. Gerace}}},\ }\href@noop {} {\bibfield  {journal}
  {\bibinfo  {journal} {Phys. Rev. A}\ }\textbf {\bibinfo {volume} {93}},\
  \bibinfo {pages} {043821} (\bibinfo {year} {2016})}\BibitemShut {NoStop}%
\bibitem [{\citenamefont {{F. Fratini and R. Ghobadi}}(2016)}]{Vienna2016}%
  \BibitemOpen
  \bibfield  {author} {\bibinfo {author} {\bibnamefont {{F. Fratini and R.
  Ghobadi}}},\ }\href@noop {} {\bibfield  {journal} {\bibinfo  {journal} {Phys.
  Rev. A}\ }\textbf {\bibinfo {volume} {93}},\ \bibinfo {pages} {023818}
  (\bibinfo {year} {2016})}\BibitemShut {NoStop}%
\bibitem [{\citenamefont {{Clemens Müller, Joshua Combes, Andrés Rosario
  Hamann, Arkady Fedorov, and Thomas M. Stace}}(2017)}]{Combes}%
  \BibitemOpen
  \bibfield  {author} {\bibinfo {author} {\bibnamefont {{Clemens Müller,
  Joshua Combes, Andrés Rosario Hamann, Arkady Fedorov, and Thomas M.
  Stace}}},\ }\href@noop {} {\bibfield  {journal} {\bibinfo  {journal} {Phys.
  Rev. A}\ }\textbf {\bibinfo {volume} {96}},\ \bibinfo {pages} {053817}
  (\bibinfo {year} {2017})}\BibitemShut {NoStop}%
\bibitem [{\citenamefont {{Y.-L. L. Fang and H. U. Baranger}}(2017)}]{Fang}%
  \BibitemOpen
  \bibfield  {author} {\bibinfo {author} {\bibnamefont {{Y.-L. L. Fang and H.
  U. Baranger}}},\ }\href@noop {} {\bibfield  {journal} {\bibinfo  {journal}
  {Phys. Rev. A}\ }\textbf {\bibinfo {volume} {96}},\ \bibinfo {pages} {013842}
  (\bibinfo {year} {2017})}\BibitemShut {NoStop}%
\bibitem [{\citenamefont {{A. R. Hamann, C. Müller, M. Jerger, M. Zanner, J.
  Combes, M. Pletyukhov, M. Weides, T. M. Stace, and A.
  Fedorov}}(2018)}]{Clemens}%
  \BibitemOpen
  \bibfield  {author} {\bibinfo {author} {\bibnamefont {{A. R. Hamann, C.
  Müller, M. Jerger, M. Zanner, J. Combes, M. Pletyukhov, M. Weides, T. M.
  Stace, and A. Fedorov}}},\ }\href@noop {} {\bibfield  {journal} {\bibinfo
  {journal} {Phys. Rev. Lett.}\ }\textbf {\bibinfo {volume} {121}},\ \bibinfo
  {pages} {123601} (\bibinfo {year} {2018})}\BibitemShut {NoStop}%
\bibitem [{\citenamefont {{C. Gonzalez-Ballestero, E. Moreno, F. J.
  Garcia-Vidal, and A. Gonzalez-Tudela}}(2016)}]{Gonzalez}%
  \BibitemOpen
  \bibfield  {author} {\bibinfo {author} {\bibnamefont {{C.
  Gonzalez-Ballestero, E. Moreno, F. J. Garcia-Vidal, and A.
  Gonzalez-Tudela}}},\ }\href@noop {} {\bibfield  {journal} {\bibinfo
  {journal} {Phys. Rev. A}\ }\textbf {\bibinfo {volume} {94}},\ \bibinfo
  {pages} {063817} (\bibinfo {year} {2016})}\BibitemShut {NoStop}%
\bibitem [{\citenamefont {{J. Ordonez-Miranda, Y. Ezzahri, and K.
  Joulain}}(2017)}]{Ordonez}%
  \BibitemOpen
  \bibfield  {author} {\bibinfo {author} {\bibnamefont {{J. Ordonez-Miranda, Y.
  Ezzahri, and K. Joulain}}},\ }\href@noop {} {\bibfield  {journal} {\bibinfo
  {journal} {Phys. Rev. E}\ }\textbf {\bibinfo {volume} {95}},\ \bibinfo
  {pages} {022128} (\bibinfo {year} {2017})}\BibitemShut {NoStop}%
\bibitem [{\citenamefont {{A. W. Snyder and J. D. Love}}(1983)}]{Snyder}%
  \BibitemOpen
  \bibfield  {author} {\bibinfo {author} {\bibnamefont {{A. W. Snyder and J. D.
  Love}}},\ }\href@noop {} {\bibfield  {journal} {\bibinfo  {journal} {Optical
  Waveguide Theory}\ } (\bibinfo {year} {Chapman \& Hall, London,
  1983})}\BibitemShut {NoStop}%
\bibitem [{\citenamefont {{A. Roulet, H. N. Le, and V.
  Scarani}}(2016)}]{Roulet}%
  \BibitemOpen
  \bibfield  {author} {\bibinfo {author} {\bibnamefont {{A. Roulet, H. N. Le,
  and V. Scarani}}},\ }\href@noop {} {\bibfield  {journal} {\bibinfo  {journal}
  {Phys. Rev. A}\ }\textbf {\bibinfo {volume} {93}},\ \bibinfo {pages} {033838}
  (\bibinfo {year} {2016})}\BibitemShut {NoStop}%
\bibitem [{\citenamefont {Scully}\ and\ \citenamefont
  {Zubairy}(1997)}]{Scully}%
  \BibitemOpen
  \bibfield  {author} {\bibinfo {author} {\bibfnamefont {M.~O.}\ \bibnamefont
  {Scully}}\ and\ \bibinfo {author} {\bibfnamefont {M.~S.}\ \bibnamefont
  {Zubairy}},\ }\href@noop {} {\emph {\bibinfo {title} {Quantum Optics}}}\
  (\bibinfo  {publisher} {Cambridge University Press},\ \bibinfo {year}
  {1997})\BibitemShut {NoStop}%
\bibitem [{\citenamefont {{K. J. Blow, Rodney Loudon, and Simon J. D.
  Phoenix}}(1990)}]{Alex90s}%
  \BibitemOpen
  \bibfield  {author} {\bibinfo {author} {\bibnamefont {{K. J. Blow, Rodney
  Loudon, and Simon J. D. Phoenix}}},\ }\href@noop {} {\bibfield  {journal}
  {\bibinfo  {journal} {Phys. Rev. A}\ }\textbf {\bibinfo {volume} {42}},\
  \bibinfo {pages} {4102} (\bibinfo {year} {1990})}\BibitemShut {NoStop}%
\bibitem [{\citenamefont {Loudon}(2000)}]{Alex90ss}%
  \BibitemOpen
  \bibfield  {author} {\bibinfo {author} {\bibfnamefont {R.}~\bibnamefont
  {Loudon}},\ }\href@noop {} {\emph {\bibinfo {title} {The quantum theory of
  light}}}\ (\bibinfo  {publisher} {Oxford University Press},\ \bibinfo {year}
  {2000})\BibitemShut {NoStop}%
\bibitem [{\citenamefont {{David Zueco, Georg M. Reuther, Sigmund Kohler, and
  Peter Hänggi}}(2009)}]{Zueco}%
  \BibitemOpen
  \bibfield  {author} {\bibinfo {author} {\bibnamefont {{David Zueco, Georg M.
  Reuther, Sigmund Kohler, and Peter Hänggi}}},\ }\href@noop {} {\bibfield
  {journal} {\bibinfo  {journal} {Phys. Rev. A}\ }\textbf {\bibinfo {volume}
  {80}},\ \bibinfo {pages} {033846} (\bibinfo {year} {2009})}\BibitemShut
  {NoStop}%
\bibitem [{\citenamefont {{T. Niemczyk, F. Deppe, H. Huebl, E. P. Menzel, F.
  Hocke, M. J. Schwarz, J. J. Garcia-Ripoll, D. Zueco, T. Hümmer, E. Solano,
  A. Marx and R. Gross}}(2010)}]{Niemczyk}%
  \BibitemOpen
  \bibfield  {author} {\bibinfo {author} {\bibnamefont {{T. Niemczyk, F. Deppe,
  H. Huebl, E. P. Menzel, F. Hocke, M. J. Schwarz, J. J. Garcia-Ripoll, D.
  Zueco, T. Hümmer, E. Solano, A. Marx and R. Gross}}},\ }\href@noop {}
  {\bibfield  {journal} {\bibinfo  {journal} {Nat. Phys.}\ }\textbf {\bibinfo
  {volume} {6}},\ \bibinfo {pages} {772} (\bibinfo {year} {2010})}\BibitemShut
  {NoStop}%
\bibitem [{\citenamefont {{Stefano Lepri and Giulio Casati}}(2011)}]{Lepri}%
  \BibitemOpen
  \bibfield  {author} {\bibinfo {author} {\bibnamefont {{Stefano Lepri and
  Giulio Casati}}},\ }\href@noop {} {\bibfield  {journal} {\bibinfo  {journal}
  {Phys. Rev. Lett.}\ }\textbf {\bibinfo {volume} {106}},\ \bibinfo {pages}
  {164101} (\bibinfo {year} {2011})}\BibitemShut {NoStop}%
\bibitem [{\citenamefont {{Eduardo Mascarenhas, Dario Gerace, Daniel Valente,
  Simone Montangero, Alexia Auffèves and M. França Santos}}(2014)}]{Eduardo}%
  \BibitemOpen
  \bibfield  {author} {\bibinfo {author} {\bibnamefont {{Eduardo Mascarenhas,
  Dario Gerace, Daniel Valente, Simone Montangero, Alexia Auffèves and M.
  França Santos}}},\ }\href@noop {} {\bibfield  {journal} {\bibinfo  {journal}
  {Europhysics Letters}\ }\textbf {\bibinfo {volume} {106}},\ \bibinfo {pages}
  {54003} (\bibinfo {year} {2014})}\BibitemShut {NoStop}%
\end{thebibliography}%


%

\end{document}